\documentclass[journal,comsoc]{IEEEtran}
\usepackage[T1]{fontenc}% optional T1 font encoding
\usepackage{amsmath}
\usepackage[cmintegrals]{newtxmath}
\usepackage{bm}
\usepackage{graphicx}
\usepackage{array}
\usepackage[caption=false,font=footnotesize]{subfig}
\usepackage{algorithm}
\usepackage{algpseudocode}
\begin{document}
\title{Amplitude-Phase-Frequency Block Modulation for OFDM-ISAC with SI-Free PAPR Reduction and Pilotless Sensing}

\author{Bensheng~Yang,~\IEEEmembership{Member,~IEEE,}
        Min~Fan,~\IEEEmembership{Member,~IEEE,}
        Haitao~Zhao,~\IEEEmembership{Senior Member,~IEEE,}
        and~Haiming~Wang,~\IEEEmembership{Member,~IEEE,}% <-this % stops a space
\thanks{Manuscript received \qquad\qquad; revised \qquad\qquad; accepted \qquad\qquad. Date of publication \qquad\qquad; date of current version \qquad. This work was supported by the Natural Science Research Start-up Foundation of Recruiting Talents of Nanjing University of Posts and Telecommunications (Grant NY224006).}%
\thanks{B. Yang and H. Zhao are with the Jiangsu Key Laboratory of Wireless Communications, Nanjing University of Posts and Telecommunications, Nanjing 210003, China (e-mail: yangbensheng@njupt.edu.cn, zhaoht@njupt.edu.cn).}%
\thanks{M. Fan is with the State Key Laboratory of Millimeter Waves, Southeast University, Nanjing 210096, China (e-mail: minfan@seu.edu.cn).}
\thanks{H. Wang is with the State Key Laboratory of Millimeter Waves, Southeast University, Nanjing 210096, China, and also with Purple Mountain Laboratories, Nanjing 211111, China (e-mail: hmwang@seu.edu.cn).}%
\thanks{Digital Object Identifier }}

\markboth{IEEE Transactions on Communications}{Yang \MakeLowercase{\textit{et al.}}: APFBM-OFDM for ISAC}

\maketitle

\begin{abstract}
Orthogonal Frequency Division Multiplexing (OFDM)-based integrated sensing and communication systems demand a unified waveform that simultaneously supports reliable data transmission, low peak-to-average power ratio (PAPR), and accurate channel sensing. 
Existing approaches multiplex communication and sensing across separate time or frequency resources, or rely on dedicated pilots for channel estimation, limiting system flexibility and increasing overhead.
This paper proposes an amplitude-phase-frequency block modulation (APFBM) scheme for OFDM that achieves waveform-level integration of communication and sensing without resource partitioning. 
Information symbols are represented on the Stokes sphere and mapped to energy-normalized Jones vectors through an unambiguous rule that establishes a deterministic phase reference per block. 
This mapping exposes a common-phase degree of freedom inherent in the signal structure.
At the transmitter, a grouped phase optimization algorithm exploits this structural freedom to reduce the PAPR without side information (SI). 
At the receiver, the same deterministic phase structure enables a Viterbi-based maximum-likelihood (ML) sequence detection algorithm that jointly recovers the optimization phases and estimates the block-wise channel amplitude and phase. 
No dedicated sensing pilots are required, as the sensing observables are extracted directly from the communication waveform. 
Closed-form error-rate and sensing-accuracy expressions are derived. 
Numerical simulations and over-the-air measurements on a software-defined radio link confirm effective PAPR reduction, accurate channel sensing, reliable phase recovery, and stable channel state information reconstruction. 
The proposed scheme trades a moderate reduction in spectral efficiency for a unified waveform design that simultaneously delivers SI-free PAPR reduction and pilotless sensing.
\end{abstract}


\IEEEpeerreviewmaketitle

\section{Introduction}
\label{sec:introduction}

\IEEEPARstart{I}{ntegrated} Sensing and Communication (ISAC) is a key enabler for sixth-generation (6G) wireless networks, aiming to reuse spectrum and hardware by sharing a common waveform for data transmission and environmental sensing \cite{Liu_ISAC_Survey_2022, Zhang_ISAC_SP_2021, 9705498}. Among candidate waveforms, Orthogonal Frequency Division Multiplexing (OFDM) is attractive owing to its maturity in communication standards and its compatibility with radar signal processing \cite{Sturm_OFDM_Radar_2011}. However, directly adopting conventional OFDM for ISAC introduces a fundamental tension among waveform efficiency, sensing accuracy, and receiver complexity\cite{MCISAC}.

First, the high peak-to-average power ratio (PAPR) inherent in OFDM forces a large power-amplifier (PA) back-off, reducing both energy efficiency and effective radiated power \cite{PAPR_Survey,jdyating}. In ISAC, this penalty is amplified: operating the PA in its nonlinear region not only degrades communication metrics such as error vector magnitude (EVM) and bit error rate (BER), but also distorts the sensing waveform, biasing parameter estimation and raising sidelobes. Consequently, PAPR reduction is not merely a communication enhancement but a system-level requirement for stable joint operation.

Second, ISAC waveforms are expected to support both robust demodulation and sensing-relevant observables such as channel and target responses. Existing approaches typically optimize one function at the expense of the other \cite{Johnston_MIMOOFDM_DFRC_2022, 9529026}, and alternative multicarrier waveforms such as orthogonal time-frequency space (OTFS) have also been explored for joint sensing and communication \cite{9557830}. For example, clipping-based PAPR reduction is simple but introduces nonlinear distortion \cite{Clipping_Distortion}. A broad range of PAPR reduction methods have been reviewed in \cite{jain2025comprehensive, sinha2023study}, where the key trade-offs involve distortion, spectral efficiency, and computational complexity.

Among distortionless schemes, selected mapping (SLM) and partial transmit sequences (PTS) \cite{SLM_Classic,lv2020genetic} are effective but require side information~(SI) for ambiguity resolution, consuming spectral resources and introducing failure modes. SI-free alternatives include active constellation extension (ACE) \cite{li2025papr}, tone injection (TI) \cite{Chen_ToneInjection_2010}, tone reservation (TR) \cite{Krongold_TR_2004}, and learning-based methods \cite{he2025mask,nguyen2024optimized}; however, none simultaneously avoids SI, eliminates waveform distortion, and supports sensing. These observations motivate the development of a distortionless, SI-free PAPR reduction mechanism that is compatible with sensing.

Third, the receiver in an OFDM-ISAC system must typically perform both communication decoding and channel/target estimation \cite{Liu_DFRC_Waveform_ICC_2022}. Classical pilot-based sensing incurs overhead and reduces data throughput, whereas pilotless or blind estimation methods often rely on iterative procedures with high complexity and potential convergence issues. An effective ISAC architecture should therefore not only reduce PAPR but also simplify joint receiver processing and minimize pilot signaling overhead.

This paper addresses the above challenges through a transceiver co-design based on \textit{amplitude-phase-frequency block modulation}~(APFBM). Inspired by the robustness of constrained block signaling \cite{Fan_APTBM_2024,FDC_Frequency,scichinaFAMMIN}, we represent information symbols on the Stokes sphere and map them to energy-normalized Jones blocks, which are then assigned to OFDM subcarrier pairs. The key insight is that the Stokes-to-Jones mapping inherently admits a \emph{common phase freedom} per block: (i)~at the transmitter, a grouped phase optimization~(GPO) scheme exploits this freedom for distortionless, SI-free PAPR reduction without resource partitioning; (ii)~at the receiver, an unambiguous mapping rule establishes a deterministic phase reference per block, enabling a non-iterative, Viterbi-based differential receiver that jointly recovers the GPO phases, demodulates data via a generalized likelihood ratio test, and extracts pilotless channel amplitude and phase observables for sensing.

The main contributions are summarized as follows:
\begin{enumerate}
    \item \textbf{APFBM-based OFDM-ISAC waveform construction:} An OFDM-compatible APFBM signal model is developed by mapping Stokes-domain information symbols to energy-normalized Jones blocks and assigning each block to a subcarrier pair. An unambiguous Stokes-to-Jones mapping rule is introduced that establishes a deterministic phase reference per block.
    \item \textbf{SI-free PAPR reduction via GPO:} A grouped GPO scheme (Algorithm~\ref{alg:gpo}) is developed in which $L_b$ consecutive APFBM blocks share a common group phase drawn from a discrete candidate set $\mathcal{P}$. The first group phase is fixed to zero as a known anchor, and the remaining $N_g - 1$ phases are optimized greedily to reduce PAPR without SI. The parameter $L_b$ provides a flexible trade-off between PAPR reduction capability and receiver complexity.
    \item \textbf{Pilotless ISAC via Viterbi-based group-phase recovery:} A non-iterative receiver is developed that recovers all group phases through differential measurements, multi-observation fusion, and Viterbi-based maximum-likelihood~(ML) sequence detection, thereby avoiding the error-propagation weakness of sequential hard-decision chains. The recovered phases enable a decision-directed maximal-ratio combining~(DD-MRC) estimator (Algorithms~\ref{alg:glrt_demod_algorithm} and~\ref{alg:dd_mrc_sensing}) that blindly estimates the intra-pair mismatch, demodulates data via a generalized likelihood ratio test~(GLRT), and applies MRC channel estimation, yielding block-wise channel estimates and mismatch-aware CSI reconstruction without sensing pilots.
    \item \textbf{USRP-based experimental validation:} Over-the-air experiments conducted with independent NI~X410 USRP devices without shared timing or reference signals verify that the proposed receiver recovers sensing-oriented CSI under practical radio channels.
\end{enumerate}

The remainder of this paper is organized as follows. Section~\ref{sec:system_model} introduces the system model and design objectives. Section~\ref{sec:bm_framework} presents the APFBM waveform construction and the proposed GPO procedure. Section~\ref{sec:receiver} describes the receiver design, the joint ISAC algorithm, and the theoretical performance analysis. Numerical and experimental results are provided in Section~\ref{sec:performance}. Section~\ref{sec:discussion} discusses the spectral-efficiency trade-off and compares the proposed scheme with existing PAPR reduction schemes. Concluding remarks are given in Section~\ref{sec:conclusion}.

\noindent\textbf{Notations:} Scalars, vectors, and sets are denoted by italic, boldface, and calligraphic letters, respectively. $N_{\text{FFT}}$ is the IFFT/FFT size, CP the cyclic prefix, $\mathcal{K}_\text{data}$ the data-subcarrier index set with $N_{\text{data}}\triangleq |\mathcal{K}_\text{data}|$ and $N_{\text{pairs}}\triangleq N_{\text{data}}/2$ APFBM blocks. $(k_{i,a},k_{i,b})$ denotes the $i$-th subcarrier pair. $L_b \geq 1$ is the shared-block size, $N_g \triangleq \lceil N_{\text{pairs}}/L_b \rceil$ the number of groups, $g(i) = \lfloor i/L_b \rfloor$ and $l(i) = i \bmod L_b$ the group index and intra-group position. $\psi_l \triangleq 2\pi l/L_b$ is the intra-group pre-phase, and $\bm{\Phi}_g=[\phi_0,\ldots,\phi_{N_g-1}]^T$ the group phase vector with entries from $\mathcal{P}$ and $\phi_0=0$ fixed. PAPR is evaluated on the CP-free OFDM symbol without oversampling.

\section{System Model}
\label{sec:system_model}

This section introduces the OFDM-based ISAC system model, including the subcarrier pairing structure, the intra-pair channel model, and the PAPR metric. Design objectives are stated in Section~\ref{ssec:design_objectives}.

\subsection{Subcarrier Pairing and Intra-Pair Channel Model}
\label{ssec:subcarrier_pairing}

\begin{figure}[!t]
    \centering
    \includegraphics[width=\columnwidth]{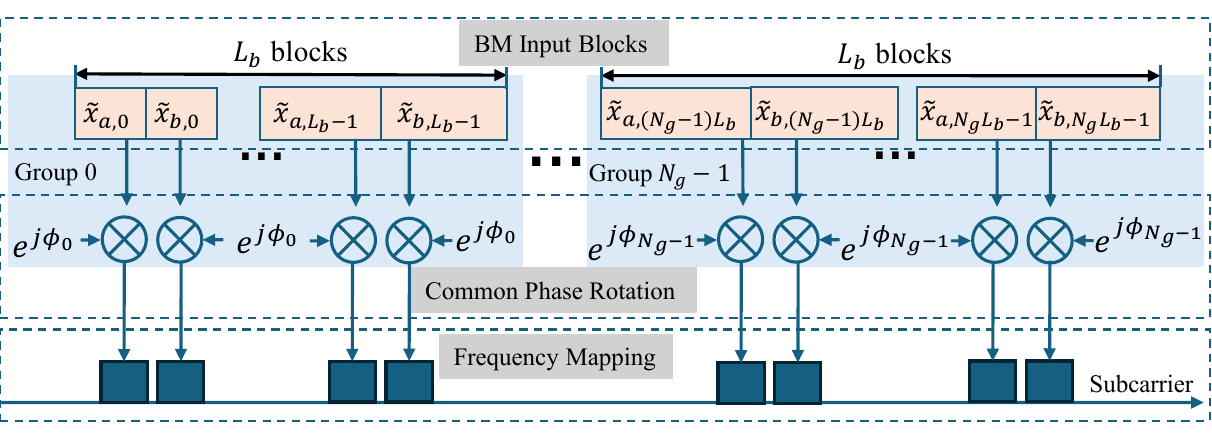}
    \caption{Data-subcarrier pairing and APFBM block mapping. Each APFBM block $\tilde{\mathbf{x}}_i=[\tilde{x}_{a,i},\tilde{x}_{b,i}]^T$ is assigned to the paired subcarriers $(k_{i,a},k_{i,b})$ and rotated by a common phase $e^{j\phi_i}$.}
    \label{fig:pairing_block_structure}
\end{figure}

Consider an OFDM-based ISAC system with $N_{\text{FFT}}$ subcarriers, of which $N_{\text{data}}$ carry data. Let $\mathcal{K}_\text{data}$ denote the set of data-subcarrier indices, with $N_{\text{data}}\triangleq|\mathcal{K}_\text{data}|$. The data subcarriers are partitioned into $N_{\text{pairs}} \triangleq N_{\text{data}}/2$ disjoint pairs, each indexed by $(k_{i,a},k_{i,b})$ for $i \in \{0,\ldots,N_{\text{pairs}}-1\}$. Each APFBM symbol is mapped onto one such subcarrier pair, as illustrated in Fig.~\ref{fig:pairing_block_structure}.

Adjacent subcarrier pairing is adopted throughout this work, i.e., $(k_{i,a},\, k_{i,b}) = (2i,\, 2i+1)$, so that the two subcarriers within each pair are separated by a single subcarrier spacing $\Delta f$. When the channel coherence bandwidth satisfies $B_c \gg \Delta f$, the two subcarriers in a pair experience nearly identical channel responses. To capture the residual intra-pair discrepancy, the channel frequency response on the second subcarrier of the $i$-th pair is modeled multiplicatively as
\begin{equation}
    H[k_{i,b}] = H[k_{i,a}]\,\Delta_i, \qquad \Delta_i \triangleq \rho_i\, e^{j\varepsilon_i},
    \label{eq:multiplicative_mismatch_model}
\end{equation}
where $\rho_i > 0$ and $\varepsilon_i \in (-\pi, \pi]$ denote the amplitude and phase mismatch, respectively. The local frequency-smoothness condition corresponds to $\rho_i \approx 1$ and $\varepsilon_i \approx 0$, the operating regime assumed hereafter. This condition is a prerequisite for the differential phase chaining receiver developed in Section~\ref{sec:receiver}.

\subsection{Frequency-Domain Received Signal Model}
\label{ssec:io_model}

After cyclic prefix (CP) removal and $N_{\text{FFT}}$-point FFT, the received symbol on the $k$-th data subcarrier is given by
\begin{equation}
    Y[k] = H[k]\, X[k] + W[k], \quad k\in\mathcal{K}_\text{data},
    \label{eq:freq_domain_io}
\end{equation}
where $X[k]$ denotes the transmitted frequency-domain symbol, $H[k]$ denotes the channel frequency response at the $k$-th subcarrier, and $W[k]\sim\mathcal{CN}(0,\sigma_w^2)$ denotes additive white Gaussian noise~(AWGN).

\subsection{PAPR Definition}
\label{ssec:papr_def}

Let $x[n]$, $n = 0, 1, \ldots, N_{\text{FFT}}-1$, denote the CP-free discrete-time OFDM waveform obtained from the $N_{\text{FFT}}$-point IFFT. The PAPR is defined as
\begin{equation}
    \mathrm{PAPR} \triangleq \frac{\displaystyle\max_{0 \le n \le N_{\text{FFT}}-1} |x[n]|^2}{\displaystyle\frac{1}{N_{\text{FFT}}}\sum_{n=0}^{N_{\text{FFT}}-1} |x[n]|^2}.
    \label{eq:papr_def}
\end{equation}

\subsection{Design Objectives}
\label{ssec:design_objectives}

The proposed scheme jointly addresses the following three objectives within a single OFDM waveform:
\begin{enumerate}
    \item \textbf{SI-free PAPR reduction:} The PAPR of the transmitted OFDM symbol is to be reduced without requiring SI at the receiver, thereby relaxing the linearity constraints on the PA.
    \item \textbf{Pilotless sensing:} Block-wise channel amplitude and phase estimation, together with smoothed data-subcarrier CSI reconstruction, are to be performed directly from the communication payload without dedicated sensing pilots.
    \item \textbf{Joint communication and sensing without resource partitioning:} Data demodulation and channel sensing are to be performed jointly from the same waveform within a single non-iterative receiver pipeline, without dedicating separate time--frequency resources to either function.
\end{enumerate}
These objectives are addressed simultaneously by the APFBM scheme developed in Sections~\ref{sec:bm_framework}--\ref{sec:receiver}.
\section{APFBM Waveform Design and Group Phase Optimization}
\label{sec:bm_framework}

This section develops the APFBM waveform construction that underlies both the SI-free PAPR reduction scheme and the pilotless receiver.
The presentation proceeds in three steps:
Section~\ref{ssec:stokes_space} introduces the Stokes-domain signal representation;
Section~\ref{ssec:unambiguous_mapping} establishes the unambiguous Jones vector mapping that resolves the common-phase ambiguity;
and Section~\ref{ssec:waveform_generation} describes the grouped phase structure and the GPO procedure.

\subsection{Signal Representation in Stokes Space}
\label{ssec:stokes_space}

In APFBM, information is encoded as coordinates on the surface of the normalized Poincar\'{e} (Stokes) sphere.
An $M$-ary APFBM constellation comprises $M$ points on this sphere, each carrying $\log_2 M$ bits and defined by a unit-norm real-valued vector
\begin{equation}
    \mathbf{S}_m = [S_{1,m},\, S_{2,m},\, S_{3,m}]^T \in \mathbb{R}^3, 
    \quad \|\mathbf{S}_m\|_2 = 1,
    \label{eq:stokes_vector}
\end{equation}
where $m \in \{1, \dots, M\}$. To maximize the minimum Euclidean distance among constellation points and thereby improve demodulation robustness, a nearly uniform point placement is adopted, e.g., via the Fibonacci lattice~\cite{Fibonacci_Lattice}.

\subsection{Unambiguous Jones Vector Mapping}
\label{ssec:unambiguous_mapping}

For transmission over a complex baseband channel, each Stokes vector $\mathbf{S}_i$ is mapped to a Jones vector $\mathbf{x}_i = [x_{a,i},\, x_{b,i}]^T \in \mathbb{C}^2$ through the relation
\begin{equation}
\left\{
\begin{aligned}
    S_{1,i} &= |x_{a,i}|^2 - |x_{b,i}|^2, \\
    S_{2,i} &= 2 \operatorname{Re}\{x_{a,i} x_{b,i}^*\}, \\
    S_{3,i} &= -2 \operatorname{Im}\{x_{a,i} x_{b,i}^*\},
\end{aligned}
\right.
\label{eq:stokes_jones_relation}
\end{equation}
subject to the unit-energy constraint $|x_{a,i}|^2 + |x_{b,i}|^2 = 1$.

A fundamental property of \eqref{eq:stokes_jones_relation} is the \emph{common-phase ambiguity}: a given Stokes vector $\mathbf{S}_i$ does not uniquely determine a Jones vector but instead specifies an equivalence class $\{e^{j\theta}\mathbf{x}_i : \theta \in [0, 2\pi)\}$.
Since the products $x_{a,i}x_{b,i}^*$, $|x_{a,i}|^2$, and $|x_{b,i}|^2$ are invariant under a common phase rotation, Stokes-domain demodulation is unaffected by this ambiguity, whereas coherent channel sensing is not.
One explicit construction is the standard half-angle parameterization,
$\mathbf{x}_i = [\cos(\vartheta_i/2),\;
\sin(\vartheta_i/2)\,e^{j\varphi_i}]^T$,
with $\vartheta_i = \arccos(S_{1,i})$ and
$\varphi_i = \arg(S_{2,i}+jS_{3,i})$,
which satisfies \eqref{eq:stokes_jones_relation}
and the energy constraint; any other valid representative is $e^{j\theta}\mathbf{x}_i$ for some $\theta$.

To resolve this ambiguity and thereby enable the differential receiver in Section~\ref{sec:receiver}, a deterministic mapping rule is adopted: $\mathbf{x}_i$ is rotated so that its first component becomes real and non-negative, yielding the \emph{unambiguous} base block
\begin{equation}
    \tilde{\mathbf{x}}_i = e^{-j \arg(x_{a,i})} \mathbf{x}_i =
    \begin{bmatrix}
        |x_{a,i}| \\[2pt] x_{b,i}\, e^{-j \arg(x_{a,i})}
    \end{bmatrix}.
    \label{eq:unambiguous_jones}
\end{equation}
For a given $\mathbf{S}_i$, this rule produces a unique $\tilde{\mathbf{x}}_i$ whose first component $\tilde{x}_{a,i} = |x_{a,i}| \in \mathbb{R}_{\ge 0}$; this deterministic phase reference is the cornerstone of the low-complexity receiver developed in Section~\ref{sec:receiver}.

\textit{Remark (South-pole singularity):} The rule in \eqref{eq:unambiguous_jones} requires $x_{a,i}\neq 0$. At the south pole $\mathbf{S}_i=[-1,\,0,\,0]^T$, the rotation is instead taken with respect to $x_{b,i}$. The Fibonacci-lattice constellation adopted in this work does not place a point at the south pole, so this degenerate case does not arise.

\subsection{Grouped GPO: Waveform Generation and Phase Structure}
\label{ssec:waveform_generation}

\subsubsection{Grouped Phase Structure}

The $N_{\text{pairs}}$ APFBM blocks are partitioned into $N_g \triangleq \lceil N_{\text{pairs}}/L_b \rceil$ groups, where $L_b \geq 1$ denotes the shared-block size. Within each group, all blocks share a common \emph{group phase} $\phi_g \in \mathcal{P}$, where $\mathcal{P}$ is a discrete candidate phase set (e.g., $\mathcal{P} = \{0,\,\pi/2,\,\pi,\,3\pi/2\}$). The $g$-th group ($g \in \{0,\ldots,N_g-1\}$) comprises blocks with indices $\{gL_b, \ldots, \min((g+1)L_b-1,\, N_{\text{pairs}}-1)\}$.

When $L_b > 1$, an additional deterministic \emph{intra-group pre-phase} $\psi_l$ is applied to the $l$-th block within each group ($l \in \{0,\ldots,L_b-1\}$) to maintain phase diversity across the grouped blocks:
\begin{equation}
    \psi_l \triangleq \frac{2\pi l}{L_b}, \quad l \in \{0, \ldots, L_b-1\}.
    \label{eq:intra_group_prephase}
\end{equation}
This sequence is fixed and known at both the transmitter and the receiver; hence, it introduces no SI. When $L_b=1$, all pre-phases reduce to zero and the grouped structure degenerates to the per-block case.

The effective phase applied to the $i$-th block is the sum of the group phase and the intra-group pre-phase:
\begin{equation}
    \phi_i = \phi_{g(i)} + \psi_{l(i)},
    \label{eq:block_phase_decomp}
\end{equation}
where $g(i) = \lfloor i/L_b \rfloor$ and $l(i) = i \bmod L_b$ denote the group index and intra-group position, respectively. The optimization variable is the group phase vector $\bm{\Phi}_g = [\phi_0, \ldots, \phi_{N_g-1}]^T$. To provide a known phase reference for the differential receiver in Section~\ref{sec:receiver}, the first group phase is fixed as an anchor:
\begin{equation}
    \phi_0 = 0.
    \label{eq:anchor_phase}
\end{equation}
This constraint removes one degree of freedom and is enforced at both ends of the link. The remaining $N_g - 1$ group phases constitute the search space of the GPO procedure described next.

\subsubsection{Frequency-Domain Symbol Construction}

\begin{figure}[!t]
    \centering
    \includegraphics[width=\columnwidth]{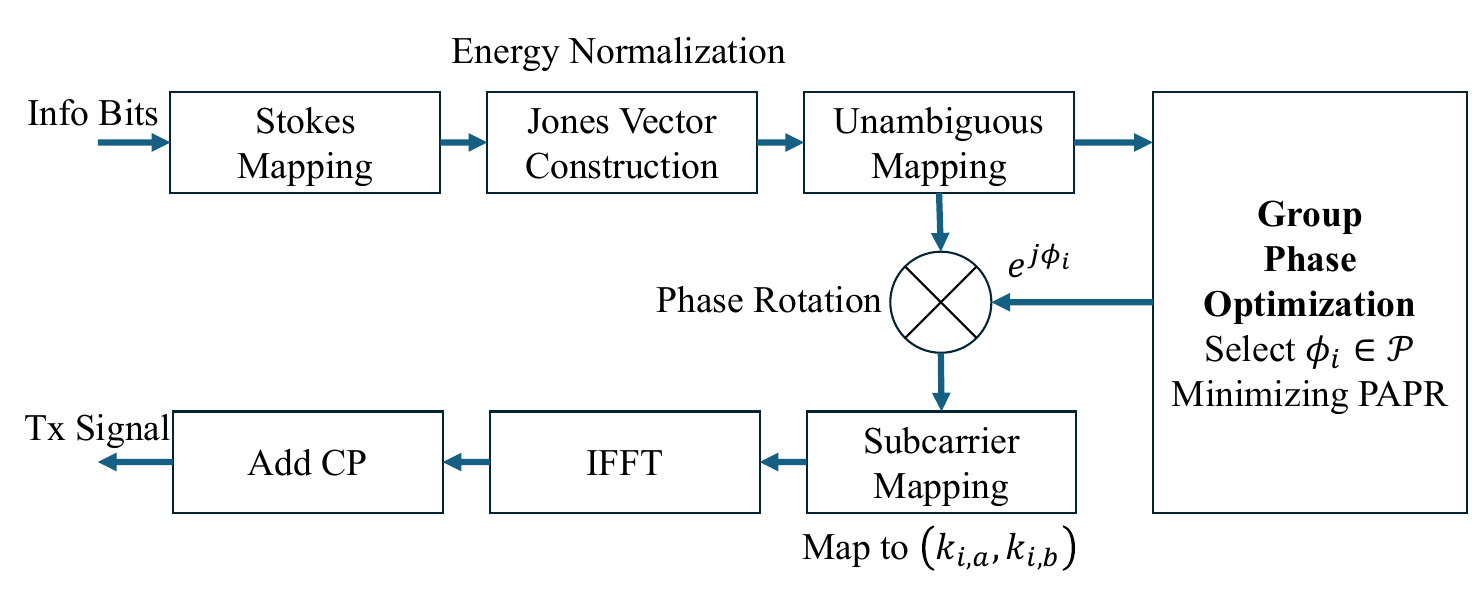}
    \caption{Transmitter processing chain for the proposed APFBM waveform with group phase optimization.}
    \label{fig:tx_processing_chain}
\end{figure}

The unambiguous base block $\tilde{\mathbf{x}}_i$ is assigned to the $i$-th subcarrier pair $(k_{i,a},\,k_{i,b})$ and rotated by the effective phase $\phi_i$, as illustrated in Fig.~\ref{fig:tx_processing_chain}. The resulting frequency-domain symbols are
\begin{align}
    X[k_{i,a}] &= e^{j\phi_i}\, \tilde{x}_{a,i} = e^{j\phi_i}\, |x_{a,i}|, \label{eq:final_symbol_a} \\
    X[k_{i,b}] &= e^{j\phi_i}\, \tilde{x}_{b,i} = e^{j\phi_i} \left(x_{b,i}\, e^{-j \arg(x_{a,i})}\right). \label{eq:final_symbol_b}
\end{align}
A key observation from \eqref{eq:final_symbol_a} is that $X[k_{i,a}]$ is a positive real amplitude scaled by $e^{j\phi_i}$; this structure is the basis of the differential phase recovery at the receiver (Section~\ref{sec:receiver}).

\subsubsection{Group Phase Optimization}

The objective of GPO is to determine the group phase vector $\bm{\Phi}_g$ (subject to $\phi_0=0$) that minimizes the PAPR of the resulting OFDM symbol without transmitting any SI to the receiver. An exhaustive search over all $|\mathcal{P}|^{N_g-1}$ candidate vectors is computationally prohibitive for practical values of $N_g$; a greedy coordinate-descent procedure is therefore adopted. Starting from a random initialization with $\phi_0=0$ fixed, the algorithm sweeps through groups $g=1,\ldots,N_g-1$ sequentially and, for each group, selects the candidate $p\in\mathcal{P}$ that yields the lowest PAPR while holding all other group phases fixed. A total of $I_{\text{g}}$ sweeps are performed, and the best group phase vector observed across all sweeps is retained as the final solution $\bm{\Phi}_g^{\star}$. The complete procedure is summarized in Algorithm~\ref{alg:gpo}.

\begin{algorithm}[!t]
\caption{Group Phase Optimization (GPO)}
\label{alg:gpo}
\begin{algorithmic}[1]
\Require Base blocks $\{\tilde{\mathbf{x}}_i\}_{i=0}^{N_{\text{pairs}}-1}$, candidate phase set $\mathcal{P}$, shared-block size $L_b$, number of sweeps $I_{\text{g}}$, IFFT size $N_{\text{FFT}}$
\Ensure Optimized group phase vector $\bm{\Phi}_g^{\star}$ and time-domain signal $\mathbf{x}^{\star}$
\State $N_g \gets \lceil N_{\text{pairs}} / L_b \rceil$
\State Compute pre-phases: $\psi_l \gets 2\pi l / L_b$, \; $l=0,\ldots,L_b-1$
\State Set $\phi_0 \gets 0$; draw $\phi_g \sim \mathrm{Uniform}(\mathcal{P})$ for $g=1,\ldots,N_g-1$
\State $\phi_i \gets \phi_{g(i)} + \psi_{l(i)}$ for all $i$ \hfill $\triangleright$ \eqref{eq:block_phase_decomp}
\State $\mathrm{PAPR}_{\min} \gets \mathrm{PAPR}(\bm{\Phi}_g)$; \; $\bm{\Phi}_g^{\star} \gets \bm{\Phi}_g$
\For{$t=1$ to $I_{\text{g}}$}
    \For{$g=1$ to $N_g-1$} \Comment{$g=0$ is the anchor}
        \State $\phi_g^{\star} \gets \phi_g$; \; $\mathrm{PAPR}_{g,\min} \gets +\infty$
        \For{each $p\in\mathcal{P}$}
            \State $\bar{\bm{\Phi}}_g \gets \bm{\Phi}_g$ with $\bar{\phi}_g \gets p$
            \State Evaluate $\mathrm{PAPR}(\bar{\bm{\Phi}}_g)$ via \eqref{eq:final_symbol_a}--\eqref{eq:final_symbol_b} and \eqref{eq:papr_def}
            \If{$\mathrm{PAPR}(\bar{\bm{\Phi}}_g) < \mathrm{PAPR}_{g,\min}$}
                \State $\mathrm{PAPR}_{g,\min} \gets \mathrm{PAPR}(\bar{\bm{\Phi}}_g)$; \; $\phi_g^{\star} \gets p$
            \EndIf
        \EndFor
        \State $\phi_g \gets \phi_g^{\star}$ \Comment{Update in place}
    \EndFor
    \If{$\mathrm{PAPR}(\bm{\Phi}_g) < \mathrm{PAPR}_{\min}$}
        \State $\mathrm{PAPR}_{\min} \gets \mathrm{PAPR}(\bm{\Phi}_g)$; \; $\bm{\Phi}_g^{\star} \gets \bm{\Phi}_g$
    \EndIf
\EndFor
\State Construct $\mathbf{x}^{\star}$ from $\bm{\Phi}_g^{\star}$ via \eqref{eq:block_phase_decomp}--\eqref{eq:final_symbol_b}
\end{algorithmic}
\end{algorithm}

\textit{Remark ($L_b$ trade-off):} A smaller $L_b$ yields more group phases ($N_g - 1$ optimization variables) and thus greater flexibility for peak suppression, at the cost of a longer differential chain and increased error-propagation risk at the receiver. Conversely, a larger $L_b$ reduces both the search space and the chain length, but restricts the degrees of freedom available for waveform shaping. The limiting case $L_b = 1$ recovers per-block GPO.

The complete OFDM symbol $\mathbf{X} \in \mathbb{C}^{N_{\text{FFT}}}$ is assembled by placing the phase-rotated blocks on the data subcarriers, zeros on the remaining subcarriers, and applying an $N_{\text{FFT}}$-point IFFT.

\section{Viterbi-Based Receiver for Joint ISAC}
\label{sec:receiver}

Building on the grouped phase structure and the unambiguous Jones vector mapping developed in Section~\ref{sec:bm_framework}, this section presents the proposed low-complexity Viterbi-based receiver.
The presentation is organized as follows:
Section~\ref{ssec:principle_differential} establishes the differential phase relation that underpins the receiver;
Section~\ref{ssec:joint_algorithm} details the three-stage joint ISAC algorithm, comprising Viterbi-based GPO group-phase recovery, GLRT-based data demodulation, and pilotless DD-MRC channel estimation;
Section~\ref{ssec:stokes_ser_analysis} derives closed-form symbol error rate (SER) expressions over Rayleigh fading channels;
Section~\ref{ssec:sensing_characterization} characterizes the sensing accuracy of the block-wise amplitude and phase estimates;
and Section~\ref{ssec:complexity_analysis} analyzes the computational complexity.

%% ---------------------------------------------------------------
\subsection{Differential Phase Chaining Principle}
\label{ssec:principle_differential}
%% ---------------------------------------------------------------

The receiver exploits a differential phase relation between the leading blocks of neighboring groups.
Recall from \eqref{eq:block_phase_decomp} that the effective phase of the $i$-th block decomposes as $\phi_i = \phi_{g(i)} + \psi_{l(i)}$, where $\phi_{g(i)} \in \mathcal{P}$ is the group phase to be recovered and $\psi_{l(i)}$ is the known intra-group pre-phase defined in \eqref{eq:intra_group_prephase}. Since $\psi_{l(i)}$ is known at the receiver, it can be subtracted, reducing the unknown quantity to the group phase~$\phi_{g(i)}$.

Consider the leading blocks of two adjacent groups $g$ and $g+1$, indexed by $i_g = g L_b$ and $i_{g+1} = (g+1)L_b$, respectively. In the noiseless case, the received first-component symbols are
\begin{align}
    Y[k_{i_g,a}]     &= h_{i_g} \, |\tilde{x}_{a,i_g}| \, e^{j(\phi_g + \psi_0)}, \label{eq:rx_leading_g}\\
    Y[k_{i_{g+1},a}] &= h_{i_{g+1}} \, |\tilde{x}_{a,i_{g+1}}| \, e^{j(\phi_{g+1} + \psi_0)}, \label{eq:rx_leading_gp1}
\end{align}
where $\psi_0 = 0$ by definition of \eqref{eq:intra_group_prephase}. Under the local phase-flatness assumption $\arg(h_{i_g}) \approx \arg(h_{i_{g+1}})$, forming the conjugate product of \eqref{eq:rx_leading_gp1} and \eqref{eq:rx_leading_g} yields
\begin{equation}
    \Delta\Psi_{\text{rx},g} \triangleq \arg\!\bigl(Y[k_{i_{g+1},a}] \cdot Y^*[k_{i_g,a}]\bigr) \approx \phi_{g+1} - \phi_g.
    \label{eq:phase_diff_key_result}
\end{equation}
The cancellation of data-dependent phase terms in \eqref{eq:phase_diff_key_result} is a direct consequence of the unambiguous mapping rule (Section~\ref{ssec:unambiguous_mapping}), which constrains $\tilde{x}_{a,i} \in \mathbb{R}_{\geq 0}$. When $L_b > 1$, each pair of adjacent groups provides $L_b$ block pairs at corresponding intra-group positions, each yielding an independent differential observation of $\phi_{g+1} - \phi_g$ after subtracting the known pre-phase difference. These $L_b$ observations are fused to improve robustness, as detailed in Section~\ref{ssec:joint_algorithm}.

\textit{Remark} (Phase-flatness condition): The relation \eqref{eq:phase_diff_key_result} requires $2L_b \Delta f \ll B_c$, where $B_c$ denotes the coherence bandwidth. For highly frequency-selective channels, a smaller $L_b$ or a non-adjacent pairing strategy may be necessary.

%% ---------------------------------------------------------------
\subsection{Three-Stage Joint ISAC Algorithm}
\label{ssec:joint_algorithm}
%% ---------------------------------------------------------------

Based on the differential phase relation established in Section~\ref{ssec:principle_differential}, a three-stage non-iterative algorithm is developed that jointly recovers the GPO group phases, demodulates data symbols, and estimates sensing-relevant channel parameters.

\subsubsection{Stage~1: Viterbi-Based GPO Group-Phase Recovery}
\label{sssec:gpo_recovery}
The first stage recovers the group phases $\{\phi_g\}_{g=0}^{N_g-1}$ via a Viterbi-based ML sequence detection procedure that avoids the error-propagation weakness of sequential greedy chaining. The procedure comprises three steps: differential observation with fusion, trellis-based sequence search, and block-phase reconstruction.

The first group phase is fixed by design (cf.\ \eqref{eq:anchor_phase}): $\hat{\phi}_0 = 0$. This anchor is enforced at both the transmitter and the receiver and requires no SI. For each adjacent group pair $(g, g+1)$, a total of $L_b$ raw differential observations are computed as
\begin{equation}
    \delta_{g,l} \triangleq \arg\!\bigl(Y[k_{i_{g+1,l},a}] \cdot Y^*[k_{i_{g,l},a}]\bigr),
    \label{eq:raw_diff_obs}
\end{equation}
where $i_{g,l} = g L_b + l$ denotes the $l$-th block in group~$g$, and the pre-phase difference cancels because $l(i_{g+1,l}) = l(i_{g,l}) = l$. When $L_b \geq 3$, the observations are fused via a trimmed mean: the extreme values are discarded and the remaining $L_b - 2$ observations are averaged to suppress outliers. For $L_b < 3$, a direct circular mean is employed instead. The resulting fused estimate of $\phi_{g+1} - \phi_g$ is denoted~$\widetilde{\Delta\Psi}_g$.

Rather than committing to a hard decision at each group boundary---which would cause a single quantization error to propagate deterministically to all subsequent groups---the $N_g - 1$ fused observations are processed jointly via a Viterbi trellis search. The trellis comprises $|\mathcal{P}|$ states per stage, one for each candidate group phase. At stage $g$ ($g = 1, \ldots, N_g - 1$), for each predecessor state $p' \in \mathcal{P}$ and current state $p \in \mathcal{P}$, the branch metric is defined as
\begin{equation}
    \lambda_g(p', p) \triangleq \left|\angle\!\left(e^{\,j\bigl(\widetilde{\Delta\Psi}_{g-1} \,-\, (p - p')\bigr)}\right)\right|^2,
    \label{eq:viterbi_branch_metric}
\end{equation}
which penalizes the squared angular distance between the observed differential phase and the hypothesized phase transition. For each state $p$ at stage $g$, the Viterbi algorithm retains only the predecessor that minimizes the cumulative path metric:
\begin{equation}
    M_g(p) = \min_{p' \in \mathcal{P}} \bigl[M_{g-1}(p') + \lambda_g(p', p)\bigr],
    \label{eq:viterbi_path_metric}
\end{equation}
with initialization $M_0(p) = 0$ for $p = \hat{\phi}_0$ and $M_0(p) = +\infty$ otherwise. After all $N_g - 1$ stages have been processed, the globally optimal group-phase sequence $\{\hat{\phi}_g\}_{g=0}^{N_g-1}$ is obtained by traceback from the terminal state with the minimum cumulative metric. The complete procedure is summarized in Algorithm~\ref{alg:viterbi_gpo}.

Because the Viterbi algorithm evaluates all $|\mathcal{P}|^2$ state transitions at every boundary and retains the globally minimum-cost path, a noisy differential observation at one boundary can be compensated by correct observations at neighboring boundaries. This global optimization eliminates the deterministic error propagation inherent in sequential hard-decision schemes.

Once all group phases $\{\hat{\phi}_g\}$ have been recovered, the effective phase of each block is reconstructed as
\begin{equation}
    \hat{\phi}_i = \hat{\phi}_{g(i)} + \psi_{l(i)},
    \label{eq:block_phase_reconstruction}
\end{equation}
where the intra-group pre-phases $\psi_{l(i)}$ are known a priori.

\begin{algorithm}[!t]
\caption{Viterbi-Based GPO Group-Phase Recovery}
\label{alg:viterbi_gpo}
\begin{algorithmic}[1]
\Require Fused observations $\{\widetilde{\Delta\Psi}_g\}_{g=0}^{N_g-2}$, candidate set $\mathcal{P}$
\Ensure Recovered group phases $\{\hat{\phi}_g\}_{g=0}^{N_g-1}$
\State Initialize $M_0(\hat{\phi}_0) \gets 0$; $M_0(p) \gets +\infty$ for $p \neq \hat{\phi}_0$
\For{$g = 1$ to $N_g - 1$} \Comment{Forward pass}
    \For{each $p \in \mathcal{P}$}
        \For{each $p' \in \mathcal{P}$}
            \State $\lambda_g(p', p) \gets |\angle(e^{j(\widetilde{\Delta\Psi}_{g-1} - (p-p'))})|^2$
        \EndFor
        \State $M_g(p) \gets \min_{p'} [M_{g-1}(p') + \lambda_g(p', p)]$
        \State $\mathrm{Surv}_g(p) \gets \arg\min_{p'} [M_{g-1}(p') + \lambda_g(p', p)]$
    \EndFor
\EndFor
\State $\hat{\phi}_{N_g-1} \gets \arg\min_p M_{N_g-1}(p)$ \Comment{Traceback}
\For{$g = N_g - 1$ down to $1$}
    \State $\hat{\phi}_{g-1} \gets \mathrm{Surv}_g(\hat{\phi}_g)$
\EndFor
\end{algorithmic}
\end{algorithm}

\subsubsection{Stage~2: GLRT-Based Data Demodulation}
\label{sssec:data_demod_glrt}
With the recovered GPO phase vector $\hat{\bm{\Phi}}$, the second stage demodulates the transmitted data symbols. This stage comprises two substeps: (i)~blind estimation of the intra-pair mismatch $\Delta_i$ from received signal statistics, and (ii)~joint symbol detection via a GLRT that exploits both Jones components together with the estimated mismatch. No dedicated pilot overhead is required.

Under the multiplicative mismatch model \eqref{eq:multiplicative_mismatch_model}, the GPO-compensated received block is defined as
\begin{equation}
    \mathbf{z}_i \triangleq e^{-j\hat{\phi}_i}
    \begin{bmatrix}
        Y[k_{i,a}] \\
        Y[k_{i,b}]
    \end{bmatrix}
    \approx a_i e^{j\theta_i}
    \begin{bmatrix}
        \tilde{x}_{a,i} \\
        \Delta_i \tilde{x}_{b,i}
    \end{bmatrix}
    + \mathbf{n}_i,
    \label{eq:gpo_compensated_block}
\end{equation}
where $a_i \triangleq |h_i|$, $\theta_i \triangleq \arg(h_i)$, $\|\tilde{\mathbf{x}}_i\|_2^2 = 1$, and $\mathbf{n}_i \sim \mathcal{CN}(\mathbf{0}, \sigma_w^2 \mathbf{I}_2)$. Let $z_{a,i}$ and $z_{b,i}$ denote the first and second components of $\mathbf{z}_i$, respectively. The demodulation procedure proceeds as follows.

\textit{Step~2-1a (Blind amplitude-mismatch estimation).}
The mismatch magnitude $\rho_i$ is estimated by exploiting the statistical relationship between the power profiles of adjacent blocks (the derivation is given in Appendix~\ref{app:blind_rho}). Defining the power sequences $P_{a,i} \triangleq |z_{a,i}|^2$ and $P_{b,i} \triangleq |z_{b,i}|^2$, the first-order differences $\Delta P_{a,i} = P_{a,i+1} - P_{a,i}$ and $\Delta P_{b,i} = P_{b,i+1} - P_{b,i}$ are computed. A sliding-window regression yields
\begin{equation}
    \hat{\beta}_i = -\frac{\sum_{j \in \mathcal{W}_i} \Delta P_{a,j} \, \Delta P_{b,j}}{\sum_{j \in \mathcal{W}_i} (\Delta P_{b,j})^2},
    \label{eq:roughness_beta}
\end{equation}
where $\mathcal{W}_i$ denotes a local window centered at block~$i$. When $\hat{\beta}_i > 0$, the amplitude mismatch is estimated as $\hat{\rho}_i = 1/\sqrt{\hat{\beta}_i}$; otherwise, $\hat{\rho}_i = 1$ is assumed. The raw estimates are then smoothed via Gaussian filtering.

\textit{Step~2-1b (Blind phase-mismatch estimation).}
The mismatch phase $\varepsilon_i$ is estimated from the cross-correlation $Y_i \triangleq z_{b,i} z_{a,i}^*$, whose argument satisfies $\arg(Y_i) \approx \varepsilon_i + \vartheta_i$, where $\vartheta_i \in \mathcal{C}_{\vartheta}$ is a data-dependent phase determined by the constellation structure. Here $\mathcal{C}_{\vartheta} \triangleq \{\arg(\tilde{x}_{b,c}\tilde{x}_{a,c}^*) : c = 1,\ldots,M\}$ denotes the set of distinct cross-product phase values induced by the $M$-point APFBM constellation. The estimation adopts an iterative quantize-and-average procedure: at each iteration, the current estimate $\hat{\varepsilon}_i$ is subtracted from $\arg(Y_i)$ and the residual is quantized to the nearest element of $\mathcal{C}_{\vartheta}$; the quantized data-dependent component is then removed and a sliding-window circular mean of the amplitude-weighted residual phasors $Y_i \, e^{-j\hat{\vartheta}_i}$ refines the mismatch estimate. Two iterations suffice for convergence. After Gaussian smoothing, the combined mismatch estimate is formed as
\begin{equation}
    \hat{\Delta}_i = \hat{\rho}_i \, e^{j\hat{\varepsilon}_i}.
    \label{eq:blind_mismatch_combined}
\end{equation}

\textit{Step~2-2 (GLRT data demodulation).}
Given the estimated mismatch, both components of each block are exploited jointly for symbol detection. The mismatch-compensated second-component observation is defined as $y_{b,i} \triangleq z_{b,i} / \hat{\Delta}_i$, and the first-component observation as $y_{a,i} \triangleq z_{a,i}$. Let $\{(\tilde{x}_{a,c},\, \tilde{x}_{b,c})\}_{c=1}^{M}$ denote the $M$-point constellation table after the unambiguous phase-factor rotation of \eqref{eq:final_symbol_a}--\eqref{eq:final_symbol_b}. The GLRT jointly detects the transmitted symbol pair as
\begin{equation}
    \hat{c}_i = \arg\max_{c \in \{1,\ldots,M\}} \;
    \bigl|\tilde{x}_{a,c}^* \, y_{a,i} + \tilde{x}_{b,c}^* \, y_{b,i}\bigr|^2,
    \label{eq:glrt_demod}
\end{equation}
where the denominator $|\tilde{x}_{a,c}|^2 + |\tilde{x}_{b,c}|^2 = 1$ is constant across all candidates and therefore omitted. The detected symbols are $\hat{x}_{a,i} = \tilde{x}_{a,\hat{c}_i}$ and $\hat{x}_{b,i} = \tilde{x}_{b,\hat{c}_i}$.

The complete Stage~2 procedure is summarized in Algorithm~\ref{alg:glrt_demod_algorithm}.

\begin{algorithm}[!t]
\caption{GLRT-Based Data Demodulation}
\label{alg:glrt_demod_algorithm}
\begin{algorithmic}[1]
\Require GPO-compensated blocks $\{(z_{a,i}, z_{b,i})\}_{i=0}^{N_{\text{pairs}}-1}$, constellation table $\{(\tilde{x}_{a,c}, \tilde{x}_{b,c})\}_{c=1}^{M}$
\Ensure Demodulated symbols $\{\hat{c}_i\}$, combined mismatch estimates $\{\hat{\Delta}_i\}$
\For{each block $i$} \Comment{Step~2-1a: Blind amplitude-mismatch estimation}
    \State Compute $\hat{\beta}_i$ from power-derivative cross-correlation \eqref{eq:roughness_beta}
    \State $\hat{\rho}_i \gets 1/\sqrt{\max(\hat{\beta}_i, \epsilon)}$; smooth across blocks
\EndFor
\For{iter $= 1$ to $2$} \Comment{Step~2-1b: Blind phase-mismatch estimation}
    \State Quantize $\arg(z_{b,i} z_{a,i}^*) - \hat{\varepsilon}_i$ to nearest $\vartheta \in \mathcal{C}_{\vartheta}$
    \State $\hat{\varepsilon}_i \gets \arg\!\bigl(\sum_{j \in \mathcal{W}_i} z_{b,j} z_{a,j}^* e^{-j\hat{\vartheta}_j}\bigr)$
\EndFor
\State $\hat{\Delta}_i \gets \hat{\rho}_i \, e^{j\hat{\varepsilon}_i}$ \Comment{Combined mismatch}
\For{each block $i$} \Comment{Step~2-2: GLRT symbol detection}
    \State $\hat{c}_i \gets \arg\max_{c}\, |\tilde{x}_{a,c}^* z_{a,i} + \tilde{x}_{b,c}^* (z_{b,i}/\hat{\Delta}_i)|^2$
    \State Extract symbols: $\hat{x}_{a,i} = \tilde{x}_{a,\hat{c}_i}$, $\hat{x}_{b,i} = \tilde{x}_{b,\hat{c}_i}$
\EndFor
\end{algorithmic}
\end{algorithm}

\subsubsection{Stage~3: Channel Sensing and Estimation}
\label{sssec:sensing_csi}
Using the demodulated symbols from Stage~2, the third stage performs decision-directed channel estimation and reconstructs a smoothed CSI profile over the data subcarriers. This stage comprises two substeps: (i)~DD-MRC channel estimation, in which the demodulated symbols serve as virtual reference signals, and (ii)~CSI reconstruction that exploits the estimated mismatch model. The output consists of block-wise amplitude and phase estimates together with a smoothed complex CSI profile over all data subcarriers, requiring no pilot overhead.

\textit{Step~3-1 (Decision-directed MRC channel estimation).}
The effective second-component reference signal is formed as $\hat{s}_{b,i} \triangleq \hat{\Delta}_i \cdot \hat{x}_{b,i}$. The channel is then estimated via maximal-ratio combining (MRC) of both observations:
\begin{equation}
    \hat{h}_i = \frac{\hat{x}_{a,i}^* \, z_{a,i} \;+\; \hat{s}_{b,i}^* \, z_{b,i}}{|\hat{x}_{a,i}|^2 + |\hat{s}_{b,i}|^2}.
    \label{eq:mrc_combination}
\end{equation}
When the GLRT decision in Stage~2 is correct and $\hat{\Delta}_i \approx \Delta_i$, the estimator \eqref{eq:mrc_combination} reduces to the minimum-variance unbiased linear estimator of $h_i$, and both Jones components contribute usable information. 

\textit{Step~3-2 (CSI reconstruction).}
The amplitude and phase of $\hat{h}_i$ are smoothed across neighboring blocks. The CSI on the first- and second-indexed data subcarriers of each pair is then reconstructed as
\begin{align}
    \hat{H}[k_{i,a}] &= \hat{h}_i, \label{eq:csi_reconstruction_a}\\
    \hat{H}[k_{i,b}] &= \hat{h}_i \cdot \hat{\Delta}_i, \label{eq:csi_reconstruction_b}
\end{align}
where the second-subcarrier CSI follows directly from the estimated mismatch model, avoiding the interpolation step required by first-component-only approaches.

The complete Stage~3 procedure is summarized in Algorithm~\ref{alg:dd_mrc_sensing}.

\begin{algorithm}[!t]
\caption{DD-MRC Sensing and CSI Reconstruction}
\label{alg:dd_mrc_sensing}
\begin{algorithmic}[1]
\Require Demodulated symbols $\{\hat{c}_i\}$, mismatch estimates $\{\hat{\Delta}_i\}$, GPO-compensated blocks $\{z_{a,i}, z_{b,i}\}$, constellation table
\Ensure Block-wise channel estimates $\{\hat{h}_i\}$, CSI profile $\{\hat{H}[k]\}$
\For{each block $i$} \Comment{Step~3-1: DD-MRC channel estimation}
    \State Extract symbols: $\hat{x}_{a,i} = \tilde{x}_{a,\hat{c}_i}$, $\hat{x}_{b,i} = \tilde{x}_{b,\hat{c}_i}$
    \State $\hat{s}_{b,i} \gets \hat{\Delta}_i \cdot \hat{x}_{b,i}$
    \State $\hat{h}_i \gets (\hat{x}_{a,i}^* z_{a,i} + \hat{s}_{b,i}^* z_{b,i}) / (|\hat{x}_{a,i}|^2 + |\hat{s}_{b,i}|^2)$
\EndFor
\State Smooth $|\hat{h}_i|$ and $\arg(\hat{h}_i)$ across blocks \Comment{Step~3-2: CSI reconstruction}
\State $\hat{H}[k_{i,a}] \gets \hat{h}_i$; \; $\hat{H}[k_{i,b}] \gets \hat{h}_i \cdot \hat{\Delta}_i$
\end{algorithmic}
\end{algorithm}

%% ---------------------------------------------------------------
\subsection{Stokes-Domain SER Analysis over Rayleigh Fading}
\label{ssec:stokes_ser_analysis}
%% ---------------------------------------------------------------

This subsection derives closed-form SER expressions that serve as analytical benchmarks for the communication performance of the proposed APFBM scheme.

For the $i$-th APFBM block on the paired subcarriers $(k_{i,a}, k_{i,b})$, the received signals under the multiplicative mismatch model \eqref{eq:multiplicative_mismatch_model} are
\begin{equation}
    y_{a,i} = h_i \, x_{a,i} + n_{a,i}, \quad y_{b,i} = h_i \, \Delta_i \, x_{b,i} + n_{b,i},
    \label{eq:rx_mismatch_pair}
\end{equation}
where $h_i \triangleq H[k_{i,a}]$. Because Stokes-domain detection is invariant to a common phase rotation, the generic Jones components $(x_{a,i}, x_{b,i})$ are used throughout this subsection without distinguishing among equivalent representatives of the same block.

To obtain a closed-form benchmark, the matched local-flat model ($\Delta_i = 1$) is adopted, under which \eqref{eq:rx_mismatch_pair} simplifies to
\begin{equation}
    y_{a,i} = h_i \, x_{a,i} + n_{a,i}, \quad y_{b,i} = h_i \, x_{b,i} + n_{b,i},
    \label{eq:rx_matched_pair}
\end{equation}
with $n_{a,i}, n_{b,i} \sim \mathcal{CN}(0, \sigma_w^2)$. Defining the normalized Stokes estimate $\hat{\mathbf{s}}_i$ from $(y_{a,i}, y_{b,i})$ as in \eqref{eq:stokes_jones_relation}, a first-order linearization at high signal-to-noise ratio (SNR) yields the equivalent AWGN model
\begin{equation}
    \hat{\mathbf{s}}_i \approx \mathbf{S}_m + \mathbf{w}_i, \quad
    \mathbf{w}_i \sim \mathcal{N}(\mathbf{0}, \sigma_s^2 \mathbf{I}_3), \quad
    \sigma_s^2 \approx \frac{2\sigma_w^2}{|h_i|^2}.
    \label{eq:stokes_awgn_bmisac}
\end{equation}
Let $d_{\min} \triangleq \min_{m \neq m'} \|\mathbf{S}_m - \mathbf{S}_{m'}\|$ denote the minimum Euclidean distance of the Stokes constellation and $N_{\mathrm{nn}}$ the average number of nearest neighbors. Applying the standard nearest-neighbor union bound, the conditional SER is approximated by
\begin{equation}
    P_s(|h_i|) \approx N_{\mathrm{nn}} \, Q\!\left(\frac{d_{\min} \, |h_i|}{2\sqrt{2\sigma_w^2}}\right).
    \label{eq:ser_nn_bmisac}
\end{equation}

Averaging over Rayleigh fading $h_i \sim \mathcal{CN}(0, \Omega)$ with $\Omega \triangleq \mathbb{E}[|h_i|^2]$, the mean SER is obtained as
\begin{equation}
    \bar{P}_s \approx \frac{N_{\mathrm{nn}}}{2}\!\left(1 - \frac{\beta\sqrt{\Omega}}{\sqrt{\beta^2 \Omega + 2}}\right), \quad
    \beta \triangleq \frac{d_{\min}}{2\sqrt{2\sigma_w^2}},
    \label{eq:ser_nn_rayleigh_bmisac}
\end{equation}
where the derivation is provided in Appendix~\ref{app:rayleigh_ser}.

%\textit{Remark} (Validity range): The expressions \eqref{eq:ser_nn_bmisac}--\eqref{eq:ser_nn_rayleigh_bmisac} are accurate for $E_b/N_0 \gtrsim 5\,\text{dB}$; at lower SNR, the nonlinear Stokes normalization introduces non-Gaussian distortion and the union bound may overestimate the true SER (cf.\ Section~\ref{ssec:joint_performance}).

%% ---------------------------------------------------------------
\subsection{Sensing Error Characterization}
\label{ssec:sensing_characterization}
%% ---------------------------------------------------------------

This subsection characterizes the estimation accuracy of the DD-MRC estimator described in Stage~3 (Section~\ref{sssec:sensing_csi}). The analysis is conducted in the high-SNR regime, where the GLRT decisions in Step~2-2 are assumed correct; only first-order terms are retained.

\subsubsection{Amplitude Estimation}
\label{sssec:amp_est}
Under the matched model ($\Delta_i = 1$) and correct GLRT demodulation, the DD-MRC estimator \eqref{eq:mrc_combination} reduces to the minimum-variance unbiased linear estimator of $h_i$ from both Jones components. The resulting complex channel estimate has variance $\sigma_w^2$, and the high-SNR amplitude error variance is (see Appendix~\ref{app:crlb})
\begin{equation}
    \operatorname{Var}(\hat{a}_i - a_i) \approx \frac{\sigma_w^2}{2},
    \label{eq:amp_variance_high_snr}
\end{equation}
consistent with the matched-model CRLB. When the GLRT produces symbol errors at lower SNR, the incorrect reference signals in Step~3-1 introduce additional estimation bias. The cross-block smoothing in Step~3-2 mitigates isolated errors, and the impact diminishes rapidly as the SNR increases.

\subsubsection{Phase Estimation}
\label{sssec:phase_est}
The DD-MRC phase estimate $\hat{\theta}_i = \arg(\hat{h}_i)$ benefits from the coherent use of both Jones components via MRC. Under the matched model and correct GLRT demodulation, a first-order expansion yields
\begin{equation}
    \operatorname{Var}(\hat{\theta}_i - \theta_i) \approx \frac{\sigma_w^2}{2 \, a_i^2},
    \label{eq:phase_variance_high_snr}
\end{equation}
%% [NOTE] The variance in \eqref{eq:phase_variance_high_snr} is $\sigma_w^2/(2a_i^2)$,
%% but the text states "equals the matched-model CRLB $\sigma_w/(\sqrt{2}a_i)$",
%% which has units of RMSE rather than variance. Verify dimensional consistency.
which corresponds to the matched-model CRLB $\sigma_w^2 / (2\,a_i^2)$ and is independent of the transmitted symbol $|\tilde{x}_{a,i}|$. This represents a significant improvement over the first-component-only estimator $\hat{\theta}_i = \arg(z_{a,i})$, whose variance $\sigma_w^2/(2a_i^2|\tilde{x}_{a,i}|^2)$ degrades severely for blocks with small $|\tilde{x}_{a,i}|$---a common occurrence on the APFBM sphere. The robustness of DD-MRC across all constellation points is attributable to the MRC step, which adaptively weights both components according to their effective SNR. At moderate-to-low SNR, GLRT demodulation errors introduce an elbow-like transition in the RMSE curve: below the threshold at which demodulation becomes reliable, the DD-MRC phase error exceeds the matched-model prediction.

\subsubsection{Cram\'{e}r--Rao Lower Bounds under Unknown Mismatch}
\label{sssec:crlb}
The matched-model results above assume perfect mismatch compensation. When $\Delta_i$ is treated as an unknown multiplicative nuisance parameter, the Fisher information available from the second component is reduced because neither $\rho_i$ nor $\varepsilon_i$ is known. The resulting single-block Cram\'{e}r--Rao lower bounds (CRLBs) become symbol-dependent:
\begin{equation}
    \operatorname{RMSE}(\hat{a}_i) \ge \frac{\sigma_w}{\sqrt{2}\,|\tilde{x}_{a,i}|}, \qquad
    \operatorname{RMSE}(\hat{\theta}_i) \ge \frac{\sigma_w}{\sqrt{2}\,a_i\,|\tilde{x}_{a,i}|},
    \label{eq:crlb_amp_phase_block}
\end{equation}
with the derivation provided in Appendix~\ref{app:crlb}. The factor $|\tilde{x}_{a,i}|$ in the denominators reflects the fact that, when $\Delta_i$ is unknown, only the first component carries exploitable information; blocks with small $|\tilde{x}_{a,i}|$ therefore exhibit higher estimation uncertainty. If $\Delta_i$ is known or $\Delta_i = 1$, \eqref{eq:crlb_amp_phase_block} reduces to the matched-model benchmarks $\sigma_w/\sqrt{2}$ and $\sigma_w/(\sqrt{2}\,a_i)$, consistent with \eqref{eq:amp_variance_high_snr} and \eqref{eq:phase_variance_high_snr}, respectively.

%% ---------------------------------------------------------------
\subsection{Computational Complexity}
\label{ssec:complexity_analysis}
%% ---------------------------------------------------------------

The proposed receiver is non-iterative and relies exclusively on linear operations. For an OFDM symbol comprising $N_{\text{pairs}}$ blocks partitioned into $N_g = \lceil N_{\text{pairs}} / L_b \rceil$ groups, the dominant computational costs are:
(i)~FFT, at $\mathcal{O}(N_{\text{FFT}} \log N_{\text{FFT}})$;
(ii)~observation fusion, where $L_b$ raw differential observations are computed and fused for each of the $N_g - 1$ group boundaries at a total cost of $\mathcal{O}(N_{\text{pairs}})$;
(iii)~Viterbi sequence search, where $N_g - 1$ trellis stages with $|\mathcal{P}|$ states each are processed at $\mathcal{O}(N_g \cdot |\mathcal{P}|^2)$; for $|\mathcal{P}| = 4$, each stage evaluates $16$ branch metrics, yielding a modest constant-factor overhead relative to the sequential quantization it replaces;
(iv)~DD-MRC sensing (Algorithm~\ref{alg:dd_mrc_sensing}), comprising blind mismatch estimation via roughness and circular mean at $\mathcal{O}(N_{\text{pairs}})$, GLRT data demodulation at $\mathcal{O}(N_{\text{pairs}} \cdot M)$ (evaluating $M$ constellation candidates per block), and MRC channel estimation at $\mathcal{O}(N_{\text{pairs}})$; and
(v)~CSI reconstruction and smoothing at $\mathcal{O}(N_{\text{pairs}})$.
The overall per-symbol complexity is $\mathcal{O}(N_{\text{FFT}} \log N_{\text{FFT}} + N_{\text{pairs}} \cdot M)$, which scales linearly in both the number of subcarriers and the constellation size. The Viterbi search adds only an $\mathcal{O}(|\mathcal{P}|^2)$ factor per trellis stage relative to the $\mathcal{O}(|\mathcal{P}|)$ of a single-pass greedy chain; for practical values $|\mathcal{P}| \in \{2, 4, 8\}$, this overhead is negligible. The resulting complexity is substantially lower than that of iterative ISAC receivers that require multiple decoding--estimation loops.

\section{Numerical and Experimental Validation}
\label{sec:performance}

This section validates the proposed APFBM--ISAC scheme through numerical simulations (Sections~\ref{ssec:setup}--\ref{ssec:numerical_results}) and over-the-air experiments on an independent USRP testbed (Sections~\ref{ssec:usrp_setup}--\ref{ssec:usrp_validation}).

%% ---------------------------------------------------------------
\subsection{Simulation Setup and Evaluation Protocols}
\label{ssec:setup}
%% ---------------------------------------------------------------

An OFDM-based ISAC system is simulated with the parameters listed in Table~\ref{tab:sim_params}. The primary modulation order is $M = 16$ (16-APFBM), corresponding to a spectral efficiency of $2$~bits per two-subcarrier block. For a fair comparison, conventional quadrature phase-shift keying (QPSK)---which achieves the same spectral efficiency of $2$~bits per complex symbol---is adopted as the baseline. The GPO search is performed over a discrete candidate set $\mathcal{P}$ for a fixed number of coordinate-descent sweeps~$I_{\text{g}}$. The channel follows the IEEE~802.11ac (TGac) Model-C delay profile, which produces a frequency-selective response across the full OFDM band while remaining approximately flat within each adjacent subcarrier pair, consistent with the per-block sensing model in Section~\ref{ssec:principle_differential}.

\begin{table}[!t]
    \renewcommand{\arraystretch}{1.2}
    \caption{Key Simulation Parameters}
    \label{tab:sim_params}
    \centering
    \begin{tabular}{l l}
        \hline
        \bfseries Parameter & \bfseries Value \\
        \hline
        OFDM FFT size ($N_{\text{FFT}}$) & 1024 \\
        Number of data subcarriers ($N_{\text{data}}$) & 896 \\
        APFBM order ($M$) & 16 \\
        Baseline scheme & QPSK (4-QAM) \\
        Phase-set cardinality ($|\mathcal{P}|$) & 4 \\
        Coordinate-descent sweeps ($I_{\text{g}}$) & 2, 4 \\
        PAPR oversampling factor ($L$) & 1 \\
        CP included in PAPR & No \\
        Channel model & TGac Model-C (frequency-selective) \\
        \hline
    \end{tabular}
\end{table}

%\textit{Communication evaluation.}
Monte Carlo simulations are conducted to validate the Stokes-domain SER expressions derived in Section~\ref{ssec:stokes_ser_analysis}. At each SNR point, APFBM symbols are drawn uniformly from the $M$-point constellation and mapped to Jones blocks via \eqref{eq:stokes_jones_relation}--\eqref{eq:unambiguous_jones}. Each block undergoes independent Rayleigh fading $h_i \sim \mathcal{CN}(0, \Omega)$ and additive noise $\mathcal{CN}(0, \sigma_w^2)$ on both components. The receiver computes the normalized Stokes estimate $\hat{\mathbf{s}}_i$ and performs nearest-neighbor detection over $\{\mathbf{S}_m\}$. The SER is obtained by averaging error counts over $10^6$~blocks per SNR point.

%\textit{Sensing evaluation.}
After GPO phase recovery, the DD-MRC estimator described in Section~\ref{sssec:sensing_csi} is applied. Blind mismatch estimation (Steps~2-1a and 2-1b of Algorithm~\ref{alg:glrt_demod_algorithm}) yields the mismatch profile $\hat{\Delta}_i$, followed by GLRT data demodulation \eqref{eq:glrt_demod}, DD-MRC channel estimation \eqref{eq:mrc_combination}, and mismatch-aware CSI reconstruction via \eqref{eq:csi_reconstruction_a}--\eqref{eq:csi_reconstruction_b}. RMSE curves are obtained by comparing the reconstructed CSI with the true channel response. To isolate the sensing estimator from GPO recovery failures, RMSE accumulation is restricted to realizations in which the recovered GPO phase sequence is correct.

%% ---------------------------------------------------------------
\subsection{Numerical Results}
\label{ssec:numerical_results}
%% ---------------------------------------------------------------

\subsubsection{PAPR Reduction}
\label{ssec:papr_performance}

\begin{figure}[!t]
    \centering
    \subfloat[Effect of shared-block size $L_b$.]{\includegraphics[width=0.49\columnwidth]{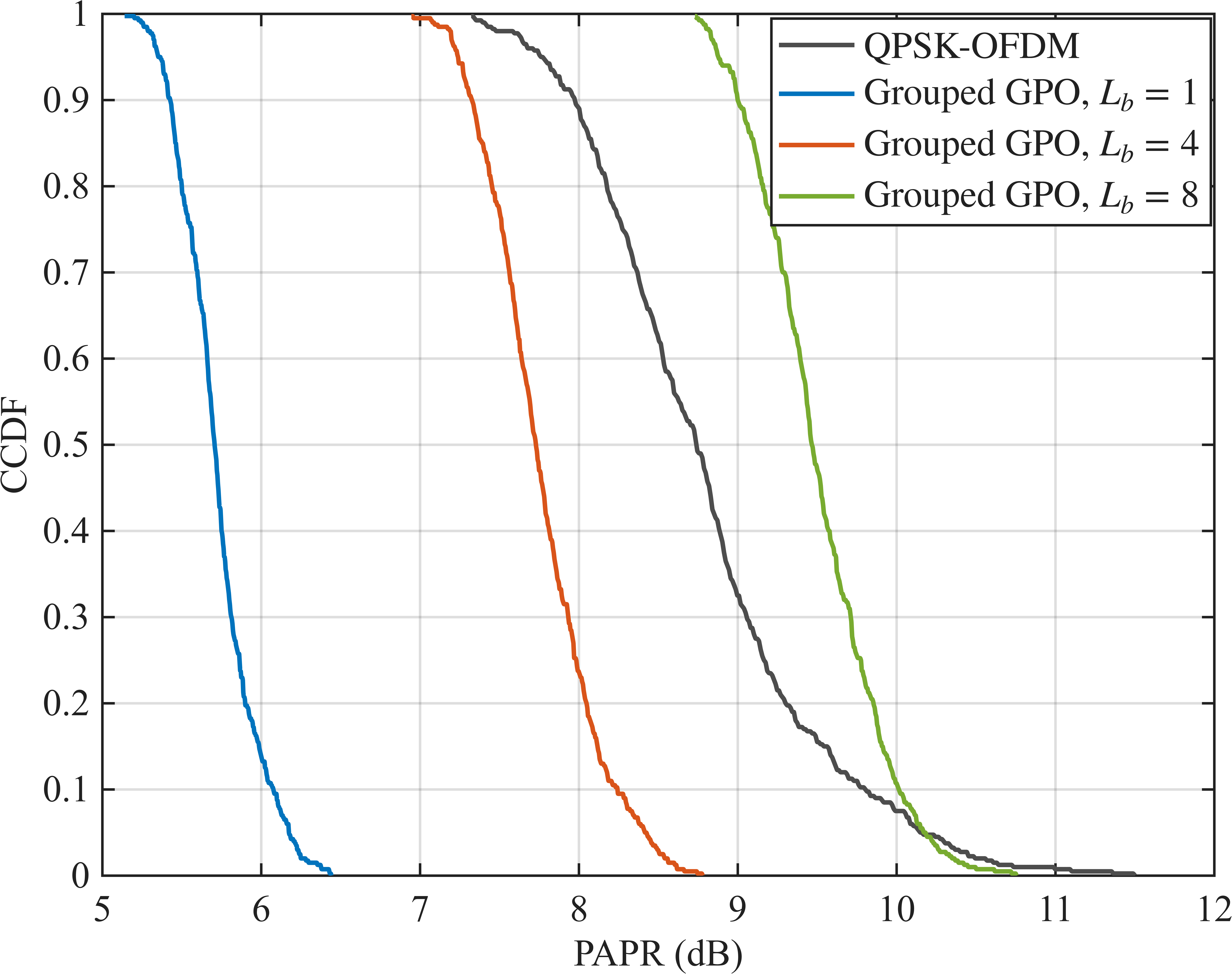}\label{fig:papr_lb}}
    \hfill
    \subfloat[Effect of phase-set cardinality $|\mathcal{P}|$ ($L_b = 1$).]{\includegraphics[width=0.49\columnwidth]{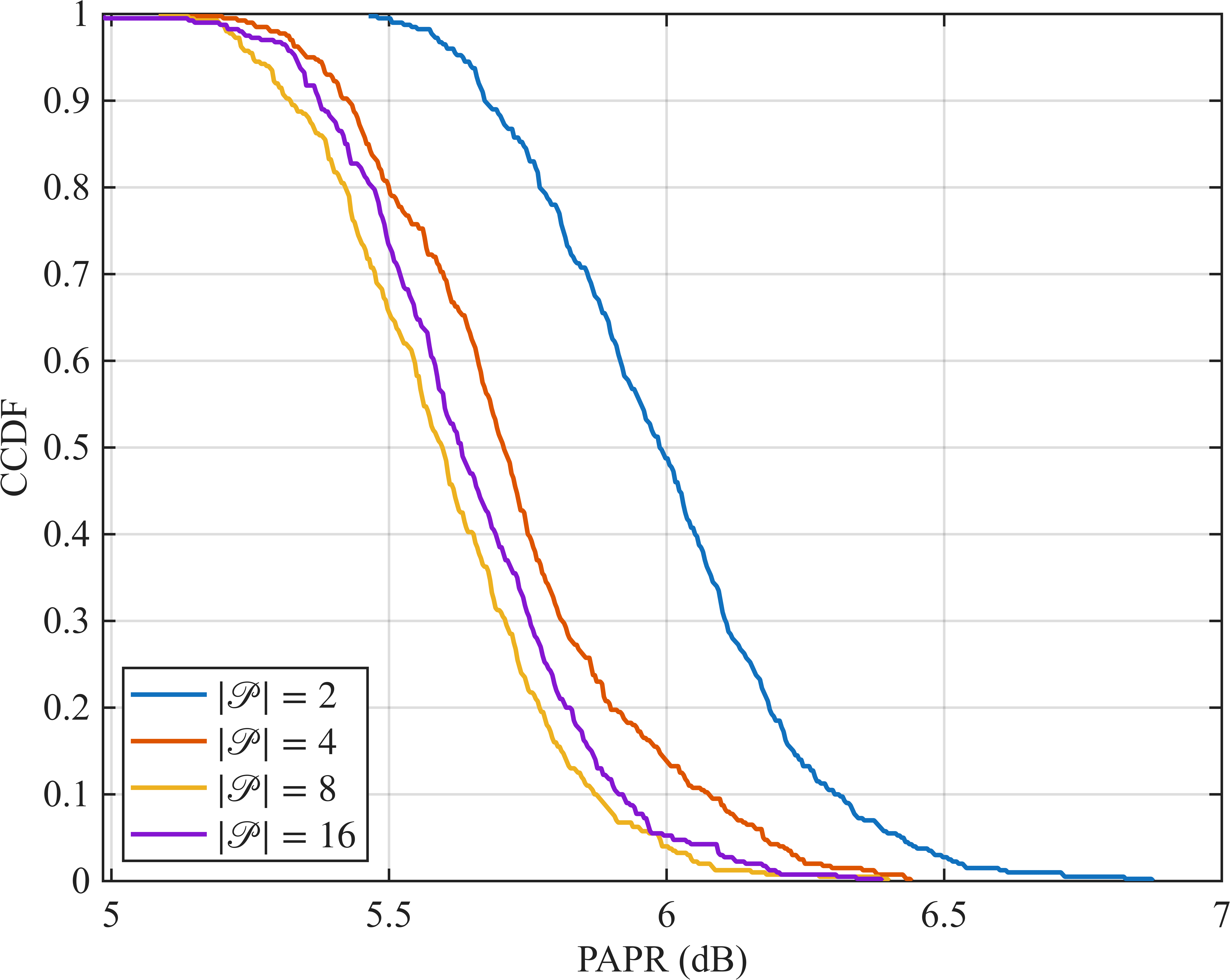}\label{fig:papr_phase_set}}
    \caption{PAPR CCDF as a function of the grouped GPO design parameters. The APFBM order is $M = 16$, yielding the same spectral efficiency as QPSK over the paired-subcarrier mapping.}
    \label{fig:papr_ccdf}
\end{figure}

Fig.~\ref{fig:papr_ccdf} examines the impact of the two principal transmitter-side design parameters: the shared-block size $L_b$ and the candidate phase-set cardinality $|\mathcal{P}|$.

In Fig.~\ref{fig:papr_ccdf}~(a), the phase set is held fixed while $L_b$ is varied. A smaller $L_b$ yields more independently adjustable phase variables and hence greater PAPR reduction, consistent with the grouped GPO design in Section~\ref{ssec:waveform_generation}. Fig.~\ref{fig:papr_ccdf}~(b) examines the complementary effect of $|\mathcal{P}|$ with $L_b = 1$: a larger $|\mathcal{P}|$ provides finer phase resolution and monotonically improves the complementary cumulative distribution function~(CCDF). Together, $L_b$ controls the number of adjustable groups, while $|\mathcal{P}|$ governs the resolution of each phase selection.

% Source: matlabcode/mainPAPR.m

\subsubsection{Communication Performance}
\label{ssec:joint_performance}

\begin{figure}[!t]
    \centering
    \subfloat[AWGN channel.]{\includegraphics[width=0.49\columnwidth]{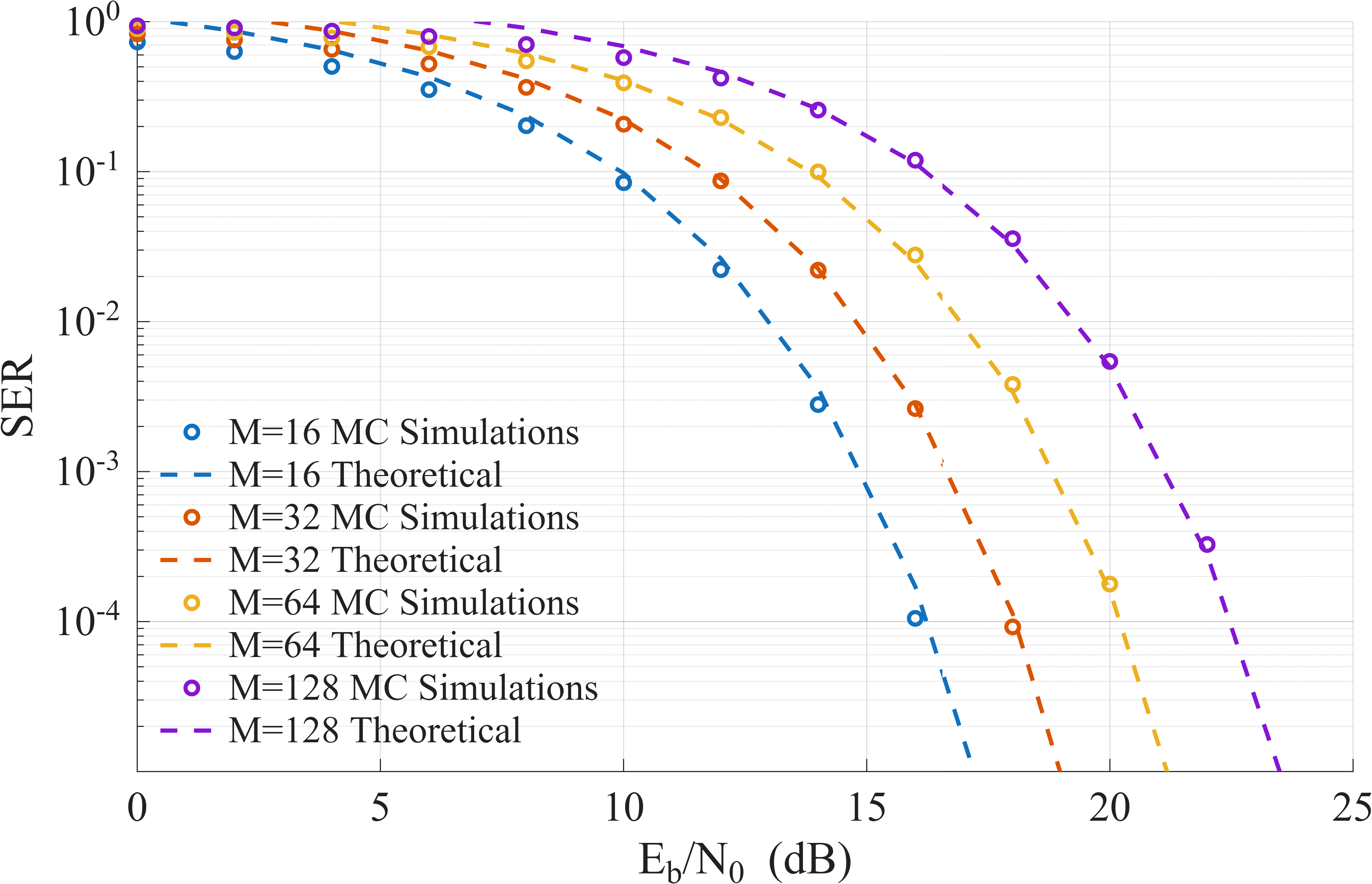}\label{fig:ser_awgn}}
    \hfill
    \subfloat[Rayleigh fading channel.]{\includegraphics[width=0.49\columnwidth]{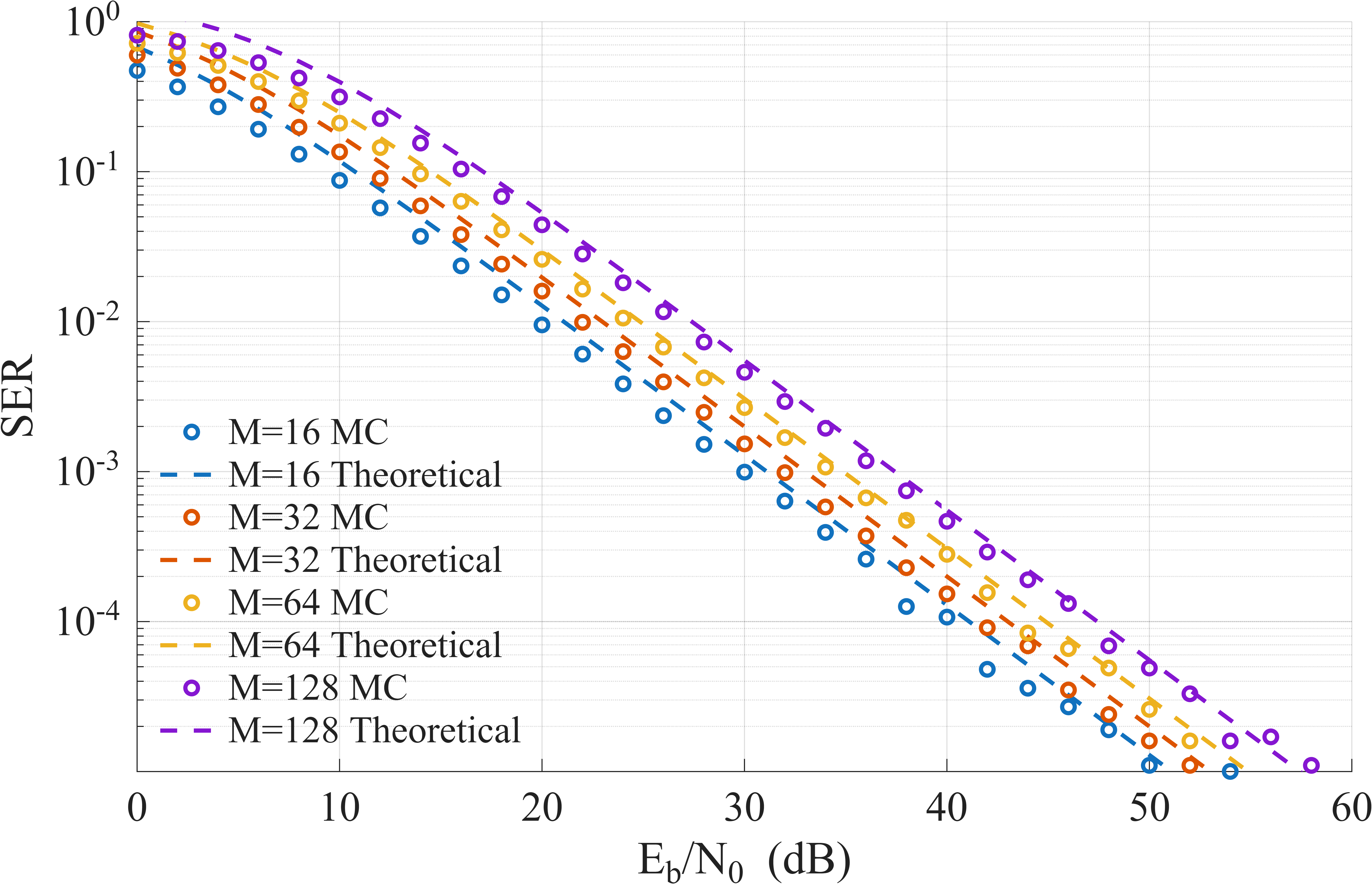}\label{fig:ser_rayleigh}}
    \caption{SER versus $E_b/N_0$ for APFBM at different modulation orders. Markers denote Monte Carlo results; dashed lines denote the nearest-neighbor approximation.}
    \label{fig:ser_performance}
\end{figure}

Fig.~\ref{fig:ser_performance} presents SER curves for APFBM constellations with $M = 16$, $32$, $64$, and $128$, comparing Monte Carlo results (markers) with the nearest-neighbor approximation \eqref{eq:ser_nn_bmisac}--\eqref{eq:ser_nn_rayleigh_bmisac} (dashed lines). In both the AWGN channel (Fig.~\ref{fig:ser_performance}~(a)) and Rayleigh fading (Fig.~\ref{fig:ser_performance}~(b)), the simulated SER closely tracks the theoretical prediction, with the fading curves correctly capturing the slope degradation relative to AWGN. The close agreement validates the analytical SER expressions in Section~\ref{ssec:stokes_ser_analysis}.

% Source: matlabcode/simulate_ber_mc.m

\subsubsection{Sensing Performance}
\label{sssec:sensing_numerical}

The pilotless sensing capability of the proposed receiver is evaluated through block-wise and reconstructed-CSI RMSE metrics. The amplitude RMSE is defined as
\begin{equation}
    \mathrm{RMSE}_{a} \triangleq \sqrt{\mathbb{E}\!\left[(\hat{a}_i - |h_i|)^2\right]},
    \label{eq:amp_rmse}
\end{equation}
and the phase RMSE is defined via the wrapped error $\varepsilon_{\theta,i} \triangleq \angle(e^{j(\hat{\theta}_i - \theta_i)})$ as
\begin{equation}
    \mathrm{RMSE}_{\theta} \triangleq \sqrt{\mathbb{E}\!\left[\varepsilon_{\theta,i}^2\right]}.
    \label{eq:phase_rmse}
\end{equation}

\begin{figure}[!t]
    \centering
    \subfloat[Block-wise amplitude RMSE.]{\includegraphics[width=0.49\columnwidth]{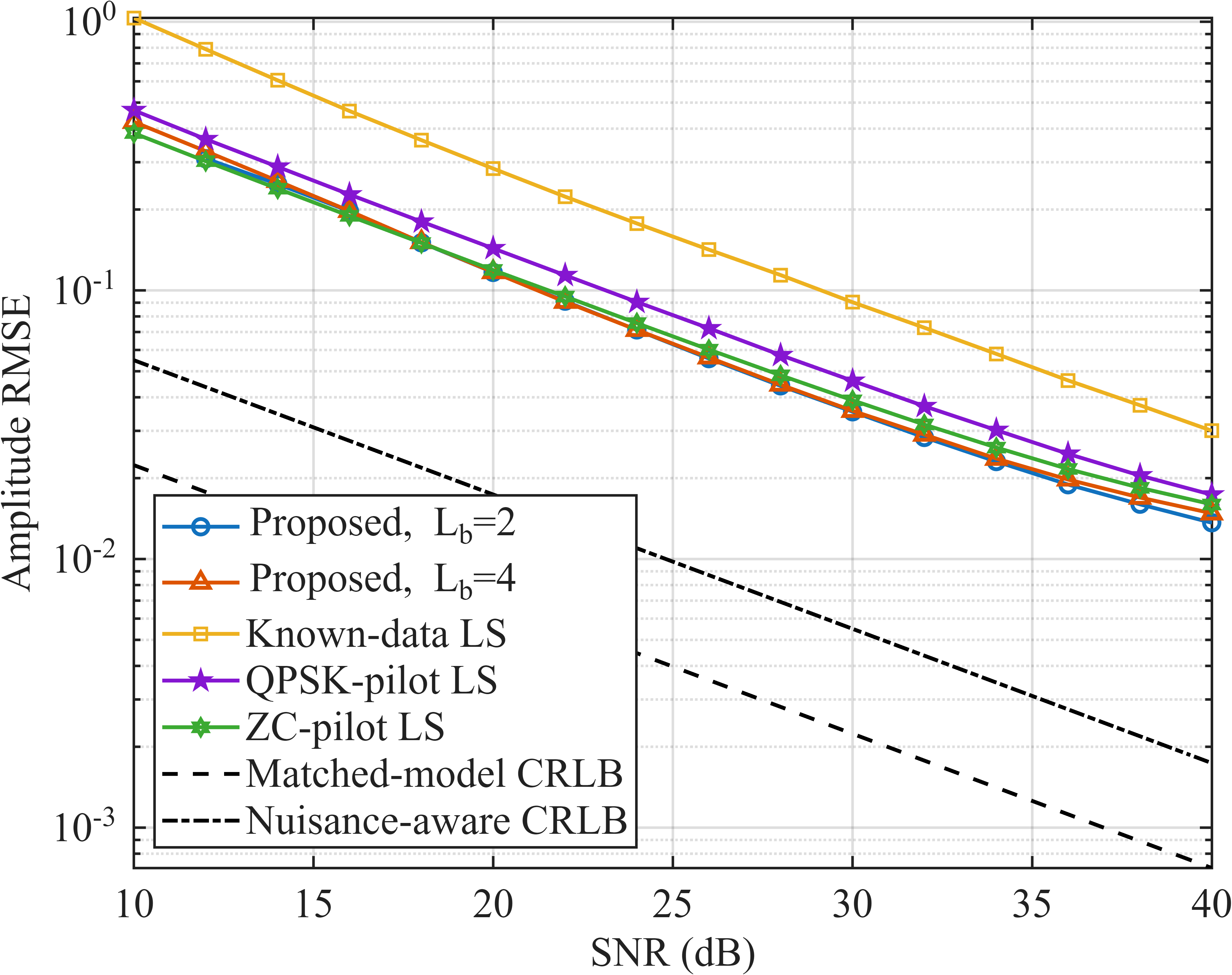}\label{fig:sensing_block_amp}}
    \hfill
    \subfloat[Block-wise phase RMSE.]{\includegraphics[width=0.49\columnwidth]{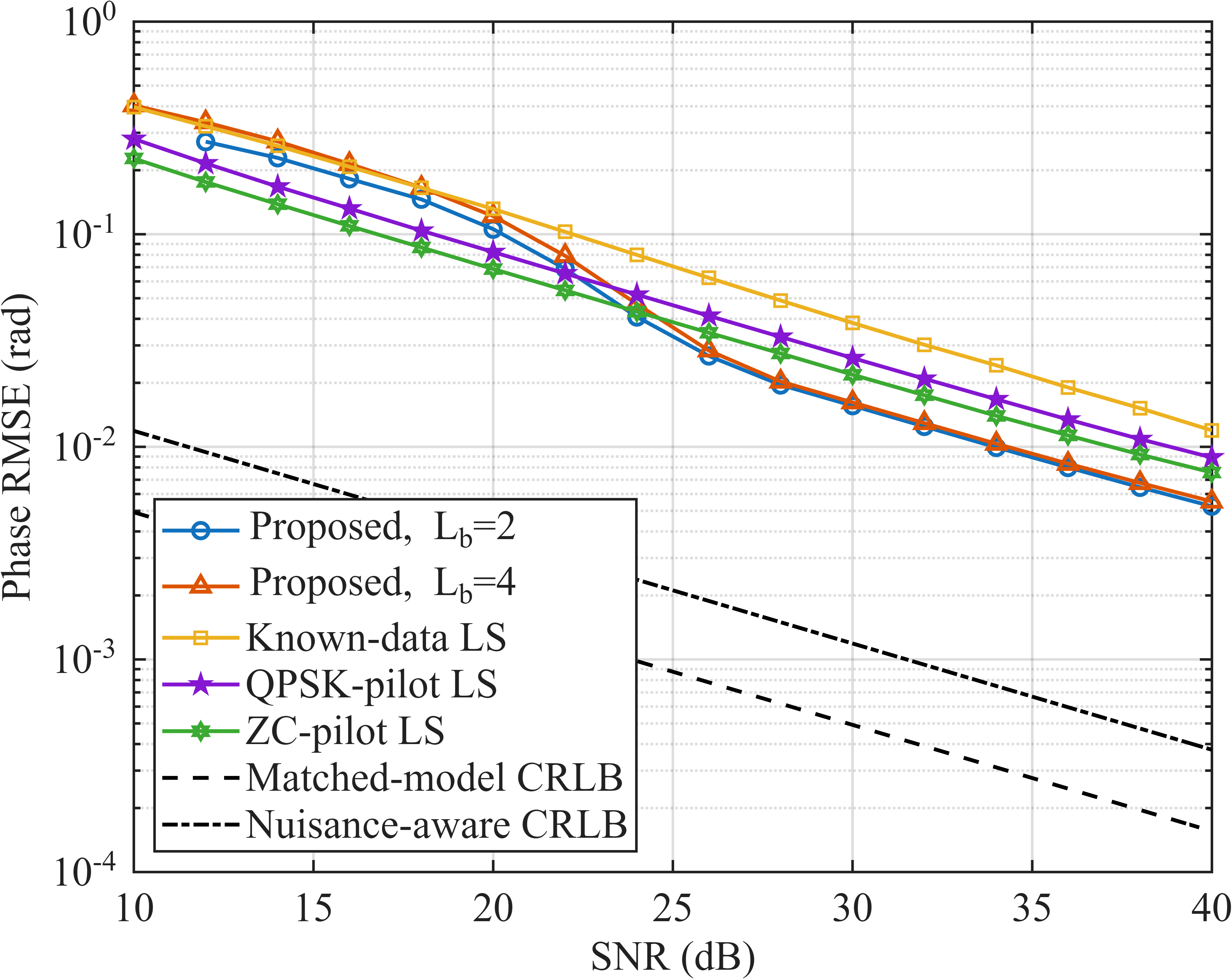}\label{fig:sensing_block_phase}}\\
    \subfloat[Reconstructed-CSI amplitude RMSE.]{\includegraphics[width=0.49\columnwidth]{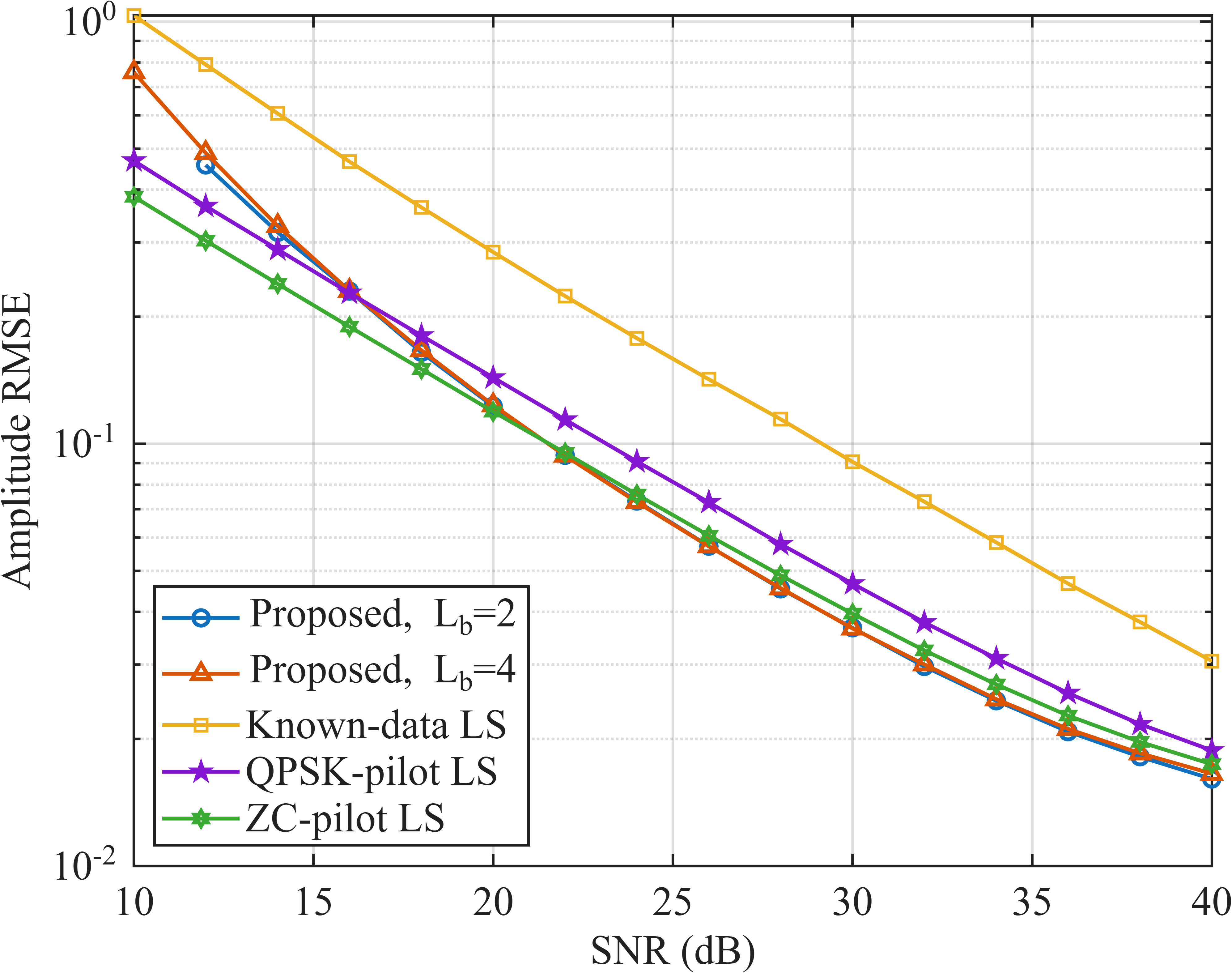}\label{fig:sensing_csi_amp}}
    \hfill
    \subfloat[Reconstructed-CSI phase RMSE.]{\includegraphics[width=0.49\columnwidth]{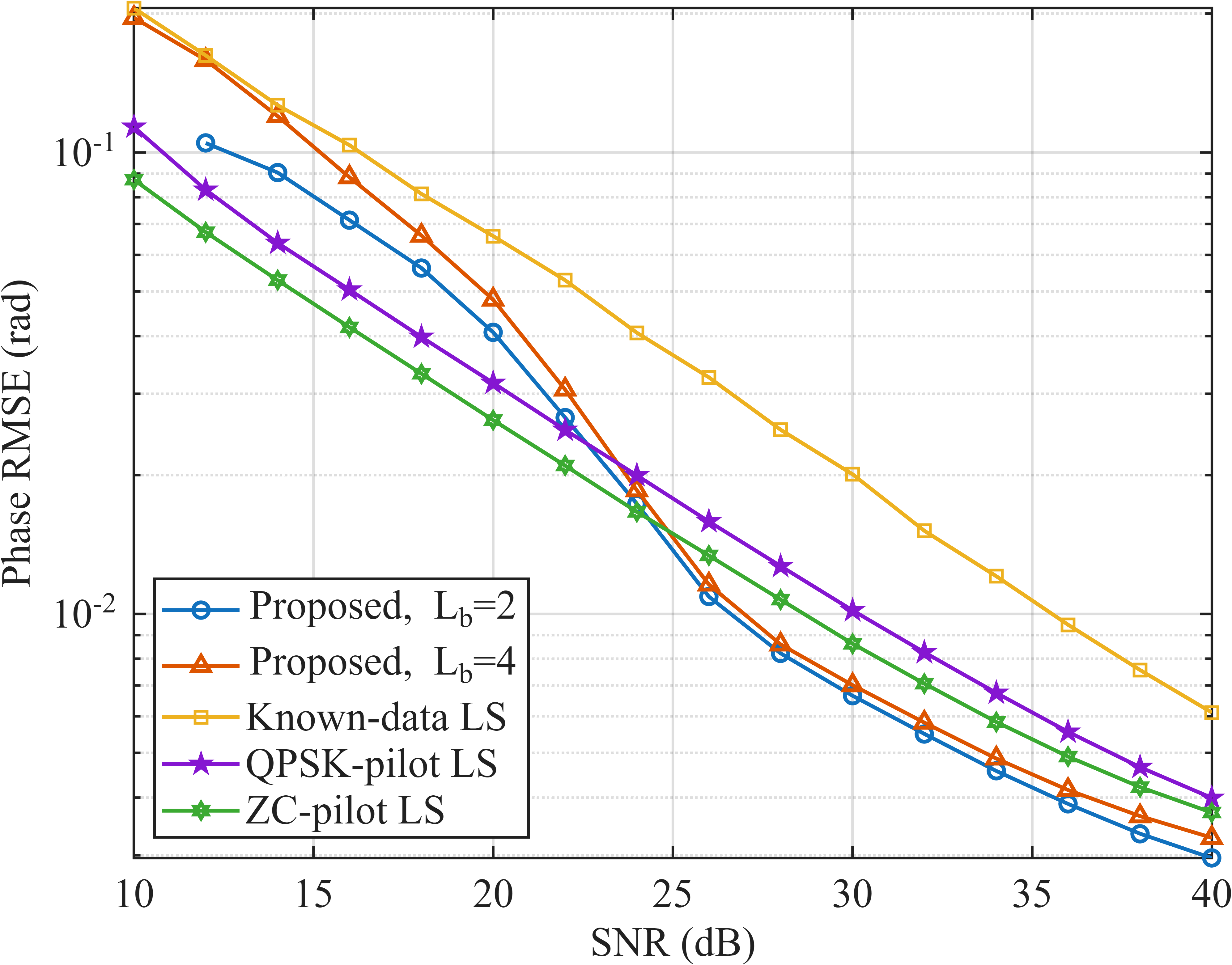}\label{fig:sensing_csi_phase}}
    \caption{Sensing RMSE versus SNR for block-wise and reconstructed-CSI channel estimation under TGac Model-C, evaluated for shared-block sizes $L_b \in \{1, 2, 4\}$ with $|\mathcal{P}| = 4$. Lower bounds correspond to the matched-model and nuisance-aware CRLBs in \eqref{eq:crlb_amp_phase_block} and Appendix~\ref{app:crlb}.}
    \label{fig:mse_performance}
\end{figure}

%\textit{Block-wise estimation.}
Fig.~\ref{fig:mse_performance} reports the sensing RMSE versus SNR under the TGac Model-C channel. The proposed DD-MRC estimator is evaluated for shared-block sizes $L_b \in \{2, 4\}$.

For block-wise amplitude estimation (Fig.~\ref{fig:mse_performance}~(a)), the proposed curves nearly overlap across $L_b$ values, confirming insensitivity to the group size. All curves remain above the matched-model and nuisance-aware CRLBs. Below approximately $18$~dB~SNR, the proposed method performs comparably to the QPSK-pilot and ZC-pilot baselines; above $18$~dB, it progressively outperforms both pilot-aided schemes.

For block-wise phase estimation (Fig.~\ref{fig:mse_performance}~(b)), the proposed curves again nearly overlap across $L_b$ values. All curves remain above the matched-model and nuisance-aware CRLBs. Below approximately $23$~dB~SNR, the proposed method exhibits higher RMSE than the QPSK-pilot and ZC-pilot baselines, attributable to residual GLRT demodulation errors. Above $23$~dB, the DD-MRC estimator surpasses both pilot-aided schemes as the GLRT demodulation becomes reliable.

\begin{figure}[!t]
    \centering
    \subfloat[$L_b$ sweep ($|\mathcal{P}|=4$).]{\includegraphics[width=0.49\columnwidth]{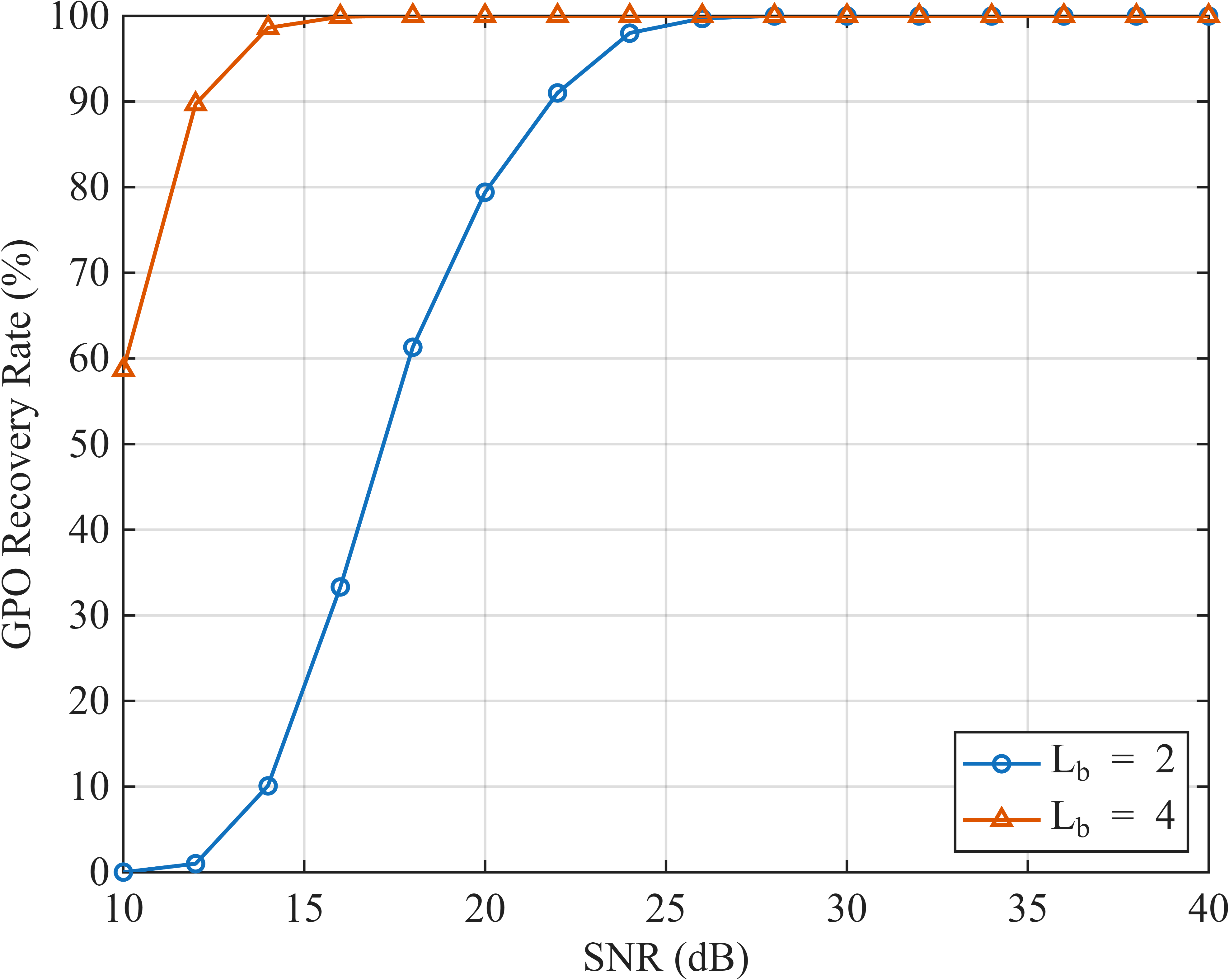}\label{fig:gpo_lb_sweep}}
    \hfill
    \subfloat[$|\mathcal{P}|$ sweep ($L_b=4$).]{\includegraphics[width=0.49\columnwidth]{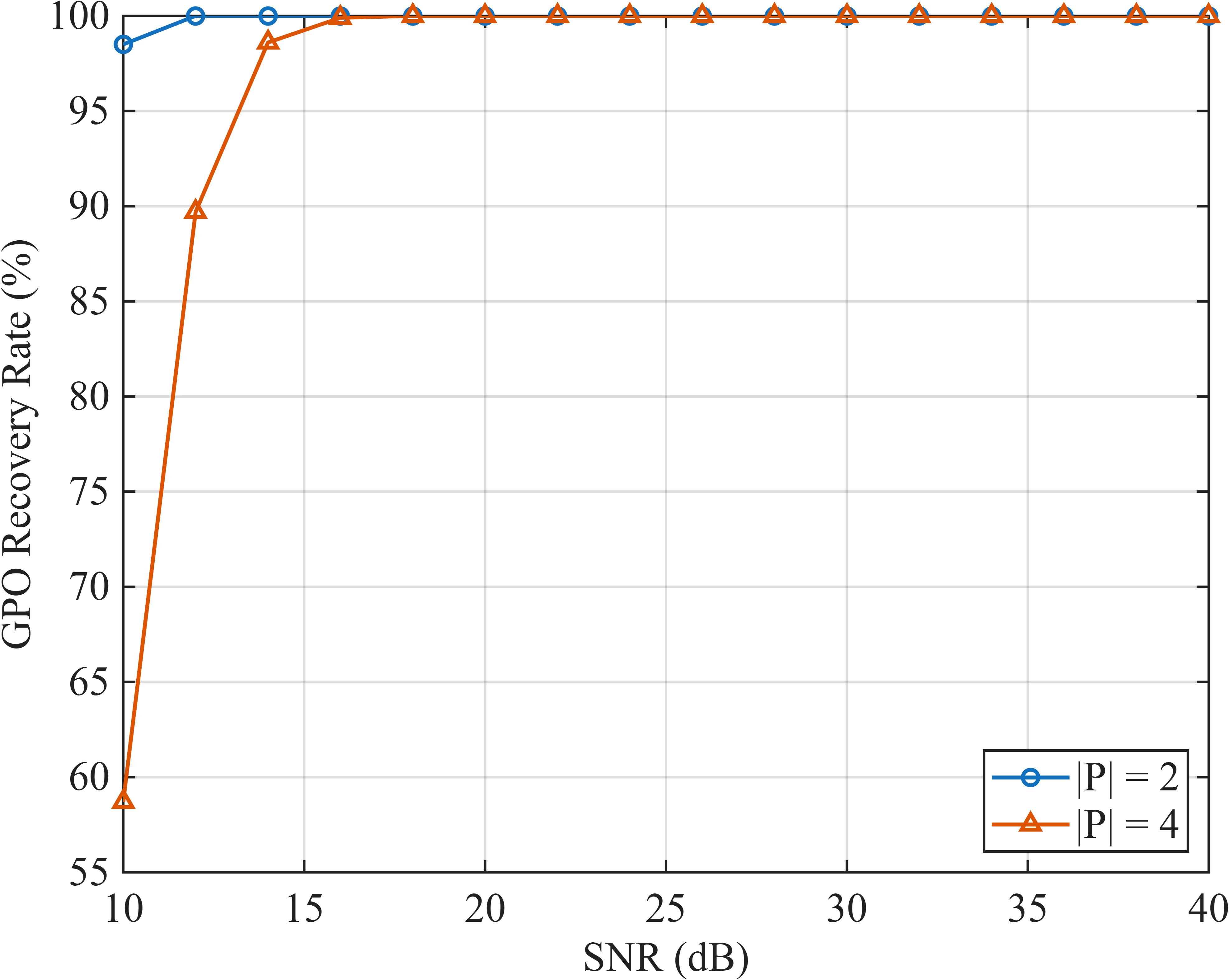}\label{fig:gpo_ps_sweep}}
    \caption{GPO recovery rate versus SNR: (a)~group size $L_b \in \{1, 2, 4\}$ with $|\mathcal{P}|=4$, and (b)~phase-set cardinality $|\mathcal{P}| \in \{2, 4\}$ with $L_b = 4$.}
    \label{fig:gpo_success_rate}
\end{figure}

%\textit{GPO recovery rate.}
The sensing RMSE is conditioned on correct GPO recovery, whose reliability is examined in Fig.~\ref{fig:gpo_success_rate}. Fig.~\ref{fig:gpo_success_rate}~(a) shows the $L_b$ sweep with $|\mathcal{P}| = 4$: $L_b = 4$ reaches $100\%$ recovery by approximately $14$~dB, while $L_b = 2$ requires approximately $24$~dB, reflecting the improved observation fusion afforded by more blocks per group boundary. Fig.~\ref{fig:gpo_success_rate}~(b) shows the $|\mathcal{P}|$ sweep with $L_b = 4$: both $|\mathcal{P}| = 2$ and~$4$ achieve near-$100\%$ recovery at modest SNR, with $|\mathcal{P}| = 2$ converging slightly faster owing to its wider decision boundary.

%\textit{Reconstructed CSI.}
The reconstructed-CSI results follow the same trends as the block-wise counterparts. For CSI amplitude (Fig.~\ref{fig:mse_performance}~(c)), the proposed method performs comparably to the pilot-aided baselines below $18$~dB and progressively outperforms them at higher SNR, with nearly identical curves across $L_b$ values. For CSI phase (Fig.~\ref{fig:mse_performance}~(d)), the proposed method exhibits higher RMSE than the pilot-aided baselines at low SNR and surpasses them above approximately $23$~dB, mirroring the block-wise phase behavior.

% Source: matlabcode/mainH.m

% Source: matlabcode/mainH.m

%% ---------------------------------------------------------------
\subsection{Over-the-Air Experimental Setup}
\label{ssec:usrp_setup}
%% ---------------------------------------------------------------

\begin{figure}[!t]
    \centering
    \includegraphics[width=0.95\columnwidth]{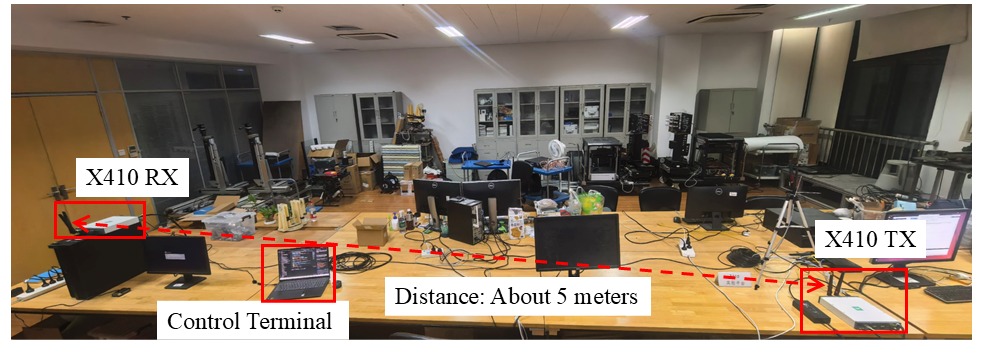}
    \caption{Photograph of the NI X410 experimental setup for over-the-air APFBM-OFDM validation. The transmitter and receiver operate without shared timing or frequency references.}
    \label{fig:x410_expt_photo}
\end{figure}

To complement the numerical evaluation, the proposed scheme is validated on a hardware testbed comprising two NI~X410 USRP devices deployed as independent transmitter and receiver over a $5$-m indoor link at $5.4$~GHz (Table~\ref{tab:usrp_params} and Fig.~\ref{fig:x410_expt_photo}). No shared reference clock, 1PPS signal, or synchronization trigger is employed; frame detection, timing alignment, and carrier-frequency offset~(CFO) compensation are performed entirely from the received waveform.

% Source: manually captured photograph (not code-generated)

The measurements are conducted in a laboratory environment with line-of-sight propagation under quasi-static conditions. The APFBM payload is transmitted without dedicated sensing pilots; the receiver relies solely on waveform-based burst synchronization, timing recovery, and CFO correction.

\begin{table}[!t]
    \renewcommand{\arraystretch}{1.2}
    \caption{Key Experimental Parameters (USRP Testbed)}
    \label{tab:usrp_params}
    \centering
    \begin{tabular}{l l}
        \hline
        \bfseries Parameter & \bfseries Value \\
        \hline
        USRP platform & NI X410 (Tx and Rx) \\
        Carrier frequency & 5.4~GHz \\
        Signal bandwidth & 61.44~MHz \\
        Sampling rate & 61.44~MHz \\
        Link distance & 5~m \\
        Propagation environment & Indoor (line-of-sight) \\
        Synchronization & Independent (no shared reference) \\
        Dedicated sensing pilots & None \\
        \hline
    \end{tabular}
\end{table}

%% ---------------------------------------------------------------
\subsection{Over-the-Air Experimental Results}
\label{ssec:usrp_validation}
%% ---------------------------------------------------------------

\subsubsection{Validation Protocol}
\label{ssec:usrp_protocol}

For each hardware capture, the transmitter emits repeated APFBM-OFDM frames with GPO. The receiver performs packet detection, frame alignment, and CFO compensation from the received waveform, after which it recovers the GPO phase sequence, estimates block-wise channel parameters, and reconstructs the smoothed CSI. Because no explicit sensing pilots are inserted, the primary validation criterion is inter-frame consistency under the same quasi-static link. A separately collected pilot-based reference is used only for offline comparison. Three metrics are reported: (i)~Stokes-domain constellation quality, (ii)~CSI consistency across repeated frames, and (iii)~GPO recovery rate.

\subsubsection{Communication Quality}
\label{sssec:usrp_comm}

\begin{figure}[!t]
    \centering
    \subfloat[Recovered 16-APFBM constellation.]{\includegraphics[width=0.48\columnwidth]{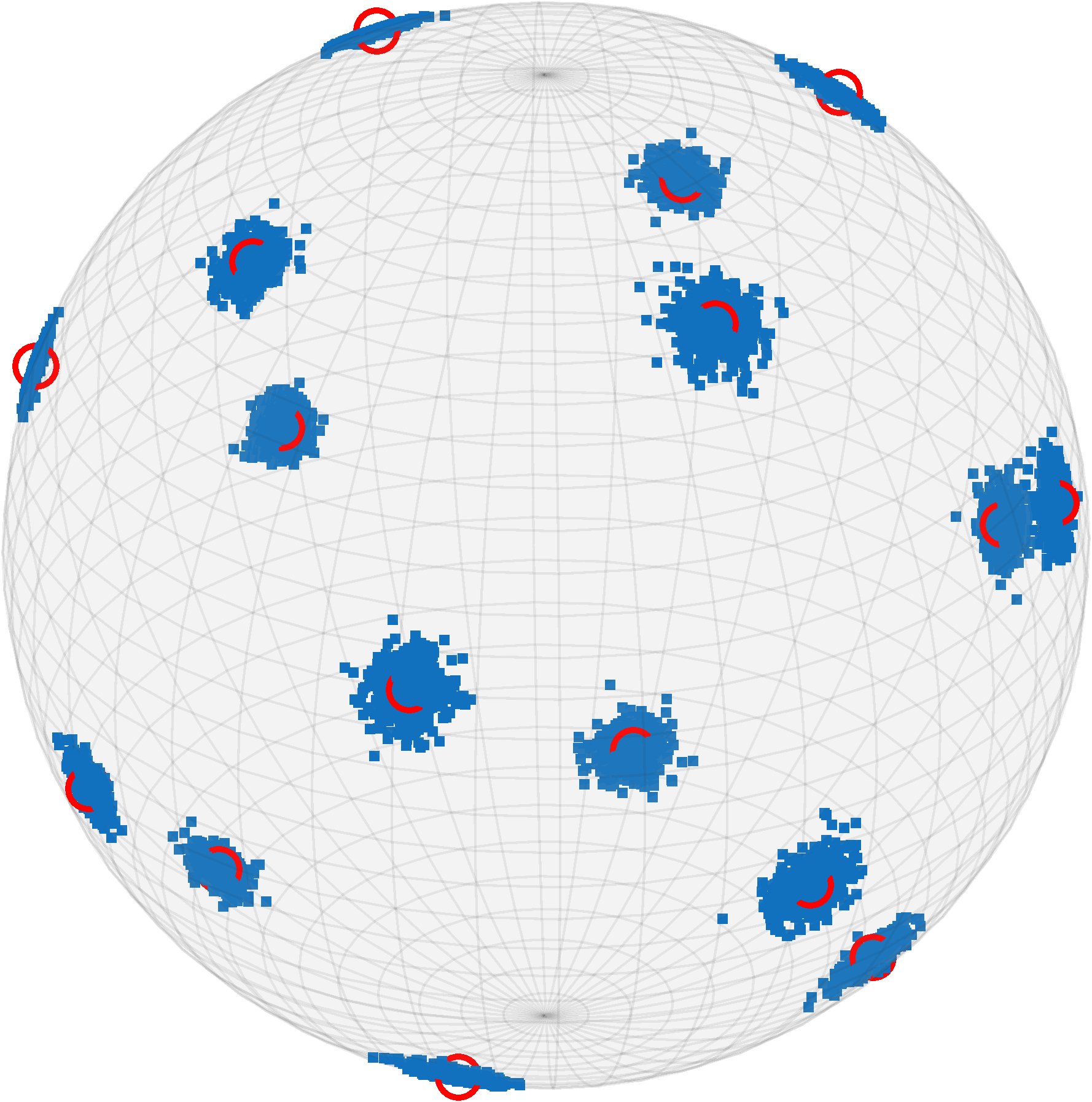}\label{fig:usrp_constellation_with_gpo_bm16}}\hfill
    \subfloat[Recovered 64-APFBM constellation.]{\includegraphics[width=0.48\columnwidth]{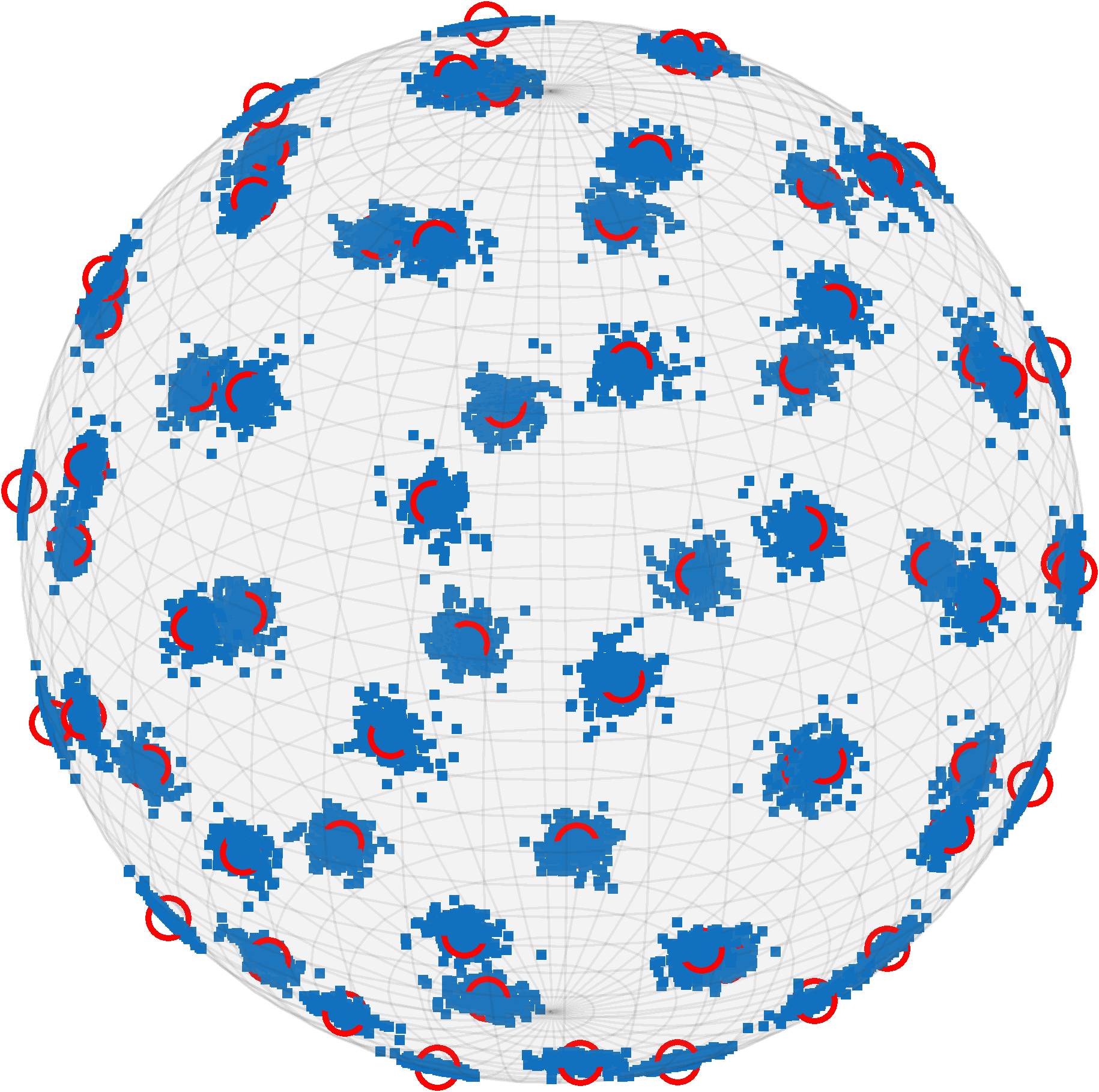}\label{fig:usrp_constellation_with_gpo_bm64}}
    \caption{Recovered Stokes-domain constellations on the 5-m indoor NI~X410 link with GPO enabled. (a)~16-APFBM: clusters are tightly grouped around the ideal points. (b)~64-APFBM: the denser constellation shows stronger spreading and partial overlap under the same hardware conditions.}
    \label{fig:usrp_comm_validation}
\end{figure}

Fig.~\ref{fig:usrp_comm_validation} presents the recovered Stokes-domain constellations. For 16-APFBM (Fig.~\ref{fig:usrp_comm_validation}~(a)), the received clusters remain tightly concentrated around the ideal points, confirming that GPO preserves the Stokes-domain symbol structure after practical RF transmission. For 64-APFBM (Fig.~\ref{fig:usrp_comm_validation}~(b)), the denser constellation exhibits stronger spreading and partial overlap, consistent with its reduced minimum Stokes distance and increased sensitivity to residual synchronization errors.

% Source: matlabcode/compare_bm16_bm64_constellations.m

\subsubsection{Sensing-Oriented CSI Consistency}
\label{sssec:usrp_sensing}

\begin{figure}[!t]
    \centering
    \subfloat[Block-wise amplitude and phase profiles.]{\includegraphics[width=0.49\columnwidth, trim=0 0 0 28bp, clip]{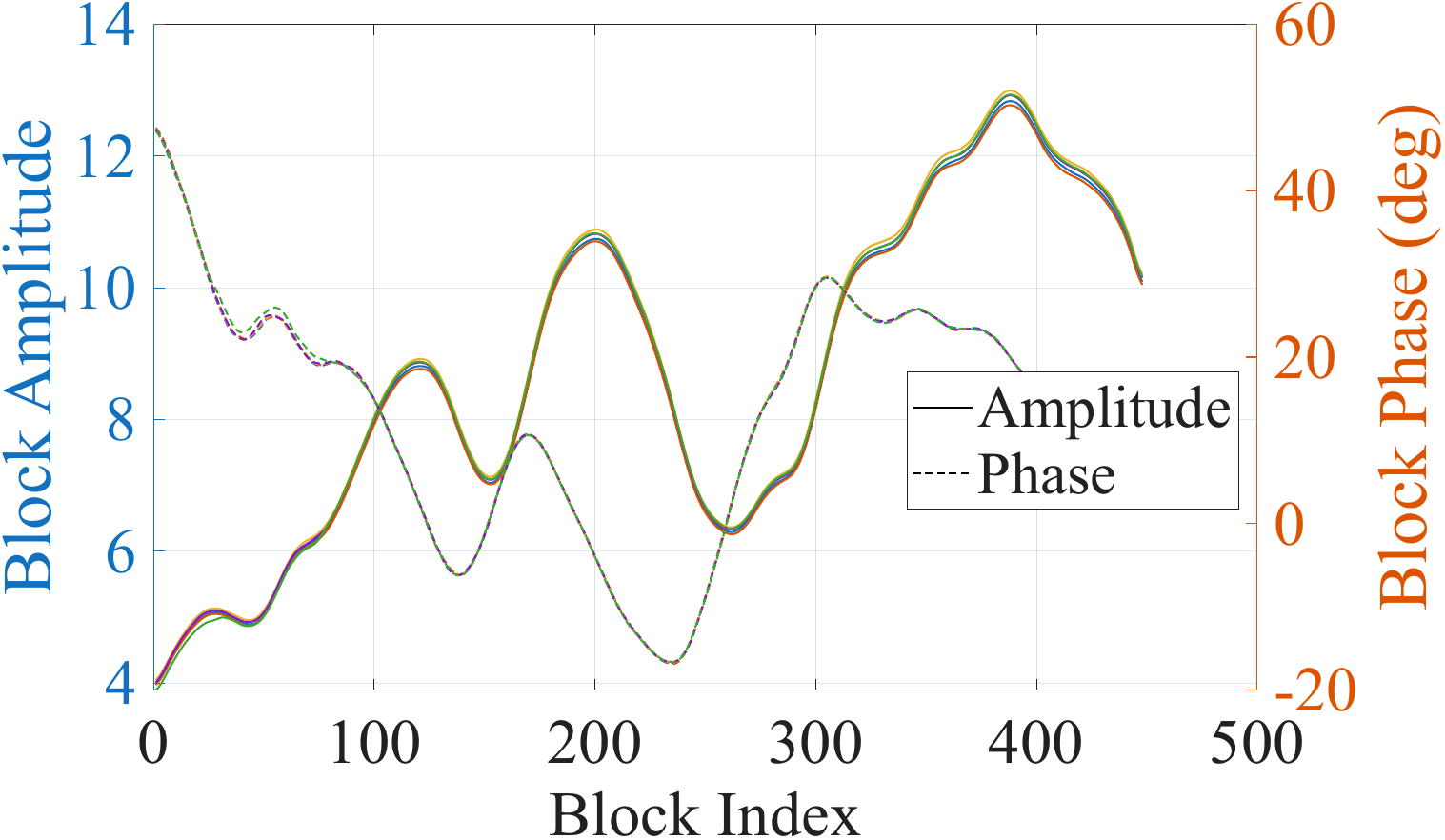}\label{fig:usrp_block_amp_phase}}\hfill
    \subfloat[Reconstructed-CSI amplitude and phase profiles.]{\includegraphics[width=0.49\columnwidth, trim=0 0 0 28bp, clip]{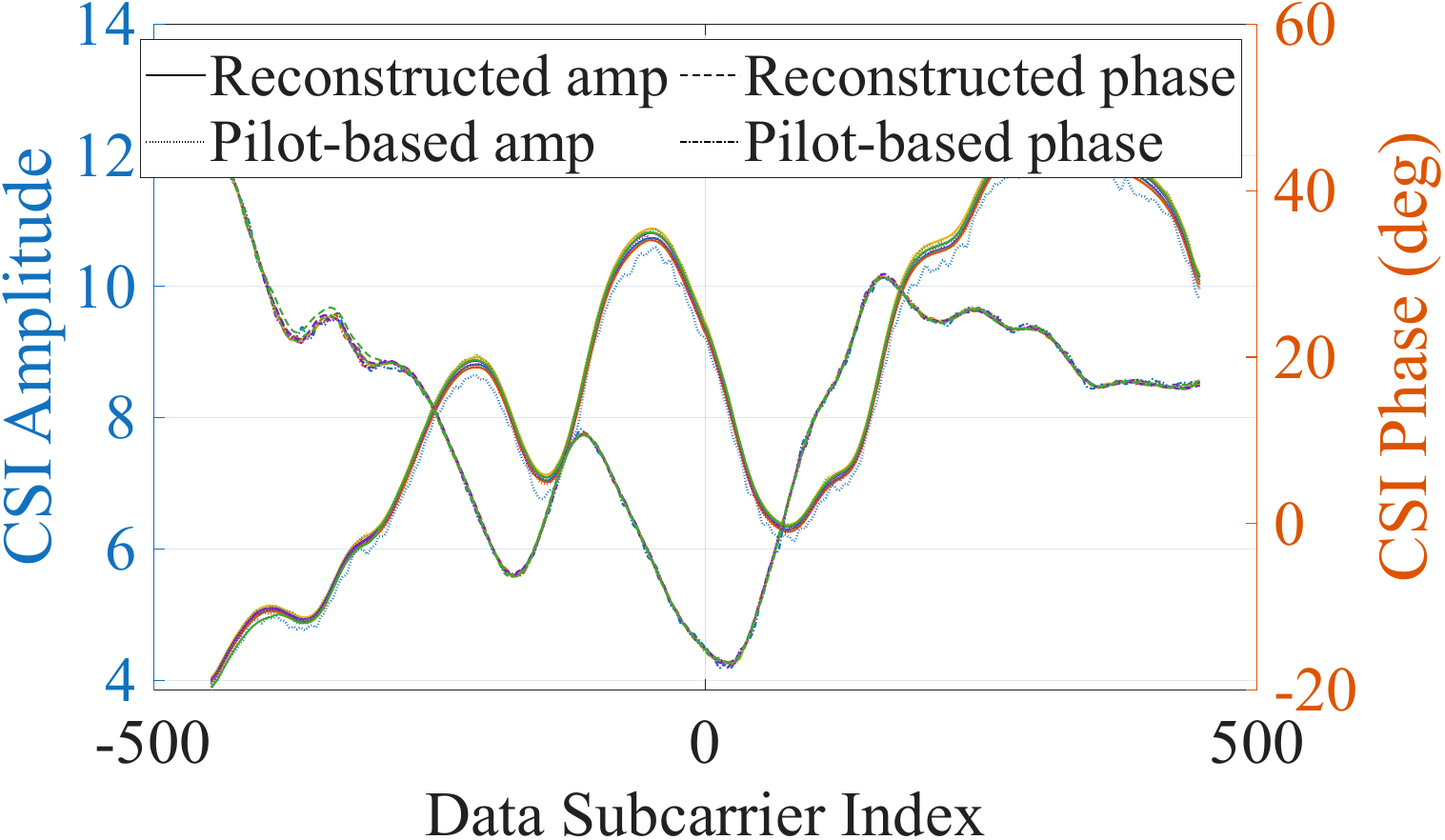}\label{fig:usrp_csi_amp_phase}}
    \caption{Sensing-oriented CSI validation on the independent NI~X410 link. Each panel overlays amplitude (left axis) and phase (right axis) profiles from several repeated frames. (a)~Block-level estimates. (b)~Reconstructed CSI compared with a pilot-based reference.}
    \label{fig:usrp_csi_validation}
\end{figure}

The central experimental result demonstrates that the proposed receiver reconstructs stable, physically consistent CSI without dedicated sensing pilots. Fig.~\ref{fig:usrp_csi_validation}~(a) overlays block-wise amplitude and phase profiles from several repeated frames, while Fig.~\ref{fig:usrp_csi_validation}~(b) shows the corresponding reconstructed-CSI profiles. In both cases, the recovered traces remain tightly clustered across captures, confirming repeatable and physically meaningful CSI extraction after waveform-based synchronization and GPO compensation.

% Source: matlabcode/generate_usrp_csi_validation_figures_multi_frame.m

\subsubsection{GPO Recovery Rate}
\label{sssec:usrp_gpo}

\begin{figure}[!t]
    \centering
    \includegraphics[width=0.75\columnwidth, trim=0 0 0 20bp, clip]{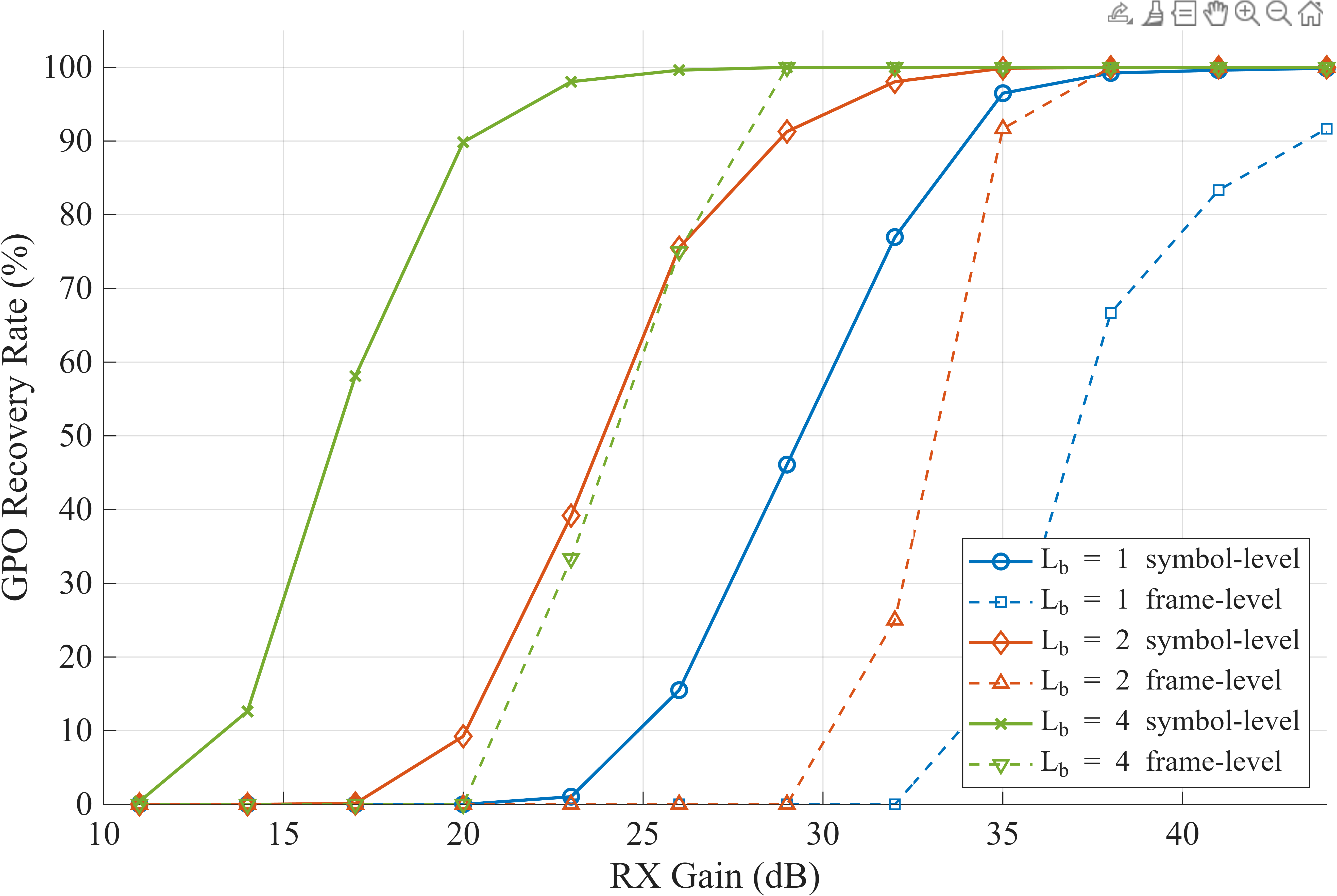}
    \caption{GPO recovery rate on the independent NI~X410 link as a function of receive gain for $L_b = 2$ and~$4$. Solid lines: symbol-level recovery rate. Dashed lines: frame-level all-correct rate.}
    \label{fig:usrp_gpo_success}
\end{figure}

Finally, the reliability of the differential phase-chaining mechanism is quantified. The receive gain is swept from $11$ to $55$~dB for $L_b = 2$ and~$4$. Fig.~\ref{fig:usrp_gpo_success} reports the symbol-level GPO recovery rate (solid) and the stricter frame-level all-correct rate (dashed). The recovery rate remains high over the practical receive-gain range for both $L_b$ settings, confirming that the observed communication quality and CSI consistency are sustained by the proposed SI-free differential phase-chaining mechanism.

% Source: matlabcode/generate_usrp_gpo_success_figure.m

\section{Discussion: Spectral Efficiency and Design Trade-offs}
\label{sec:discussion}

The APFBM scheme improves PAPR performance and simplifies receiver processing at the cost of reduced spectral efficiency~(SE). This section quantifies the resulting design trade-offs. Table~\ref{tab:tradeoff} first compares a conventional QAM-based system with the proposed APFBM-based system across five design dimensions. An effective-throughput metric that accounts for PA back-off is then introduced, followed by a comparison with other SI-free PAPR reduction schemes.

\begin{table}[!t]
    \renewcommand{\arraystretch}{1.2}
    \caption{Trade-off Comparison: Conventional QAM vs.\ Proposed APFBM}
    \label{tab:tradeoff}
    \centering
    \begin{tabular}{l l l}
        \hline
        \bfseries Feature & \bfseries Conventional QAM & \bfseries Proposed APFBM \\
        \hline
        Spectral efficiency & $\log_2(M)$ b/s/Hz & $\log_2(M)/2$ b/s/Hz \\
        PAPR & High & Low (SI-free GPO) \\
        PA requirement & High linearity & Relaxed \\
        Comm.\ receiver & Linear equalization & No phase tracking \\
        Sensing receiver & Pilot-aided & Pilotless, non-iterative \\
        \hline
    \end{tabular}
\end{table}

As shown in Table~\ref{tab:tradeoff}, the two-subcarrier-per-block mapping halves the SE relative to conventional QAM. In return, the PAPR is substantially reduced without SI (Section~\ref{ssec:waveform_generation}), the communication receiver avoids explicit carrier-phase tracking (Section~\ref{ssec:joint_algorithm}), and the sensing receiver extracts per-subcarrier CSI without dedicated pilots (Section~\ref{ssec:principle_differential}).

\subsection{Effective Throughput Under PA Back-Off}
A fair comparison between APFBM and conventional QAM should account for the PA operating point. For a Class-A or Class-AB amplifier, the achievable average transmit power is limited by the required input back-off~(IBO), which scales with the signal PAPR. Let $P_{\text{sat}}$ denote the PA saturation power and $\text{PAPR}_\epsilon$ the PAPR (in linear scale) at a target CCDF level~$\epsilon$. The achievable average transmit power is then approximately $P_{\text{avg}} \approx P_{\text{sat}}/\text{PAPR}_\epsilon$, and the effective throughput can be expressed as
\begin{equation}
    T_{\text{eff}} = \eta_{\text{SE}} \cdot \log_2(M) \cdot (1 - \bar{P}_s) \cdot \frac{P_{\text{sat}}}{\text{PAPR}_\epsilon},
    \label{eq:effective_throughput}
\end{equation}
where $\eta_{\text{SE}} = 1$ for QAM and $\eta_{\text{SE}} = 1/2$ for APFBM, and $\bar{P}_s$ denotes the average SER. Although APFBM incurs a factor-of-two SE penalty, its lower $\text{PAPR}_\epsilon$ allows the PA to operate closer to saturation, partially compensating the SE loss. For example, a $3$\,dB GPO-induced PAPR reduction (cf.\ Fig.~\ref{fig:papr_ccdf}) doubles the achievable average power, recovering half of the SE deficit in link budget.

\subsection{Comparison with Other SI-Free PAPR Reduction Schemes}
Table~\ref{tab:tradeoff_extended} compares the SE overhead and additional capabilities of several representative PAPR reduction families.

\begin{table}[!t]
    \renewcommand{\arraystretch}{1.2}
    \caption{SE Overhead and Capabilities of PAPR Reduction Schemes}
    \label{tab:tradeoff_extended}
    \centering
    \begin{tabular}{l c c c}
        \hline
        \bfseries Scheme & \bfseries SE loss & \bfseries SI-free & \bfseries Sensing \\
        \hline
        ACE  & $0\%$ & Yes & No \\
        TR   & $5$--$10\%$ & Yes & No \\
        SLM/PTS & $<5\%$ & No & No \\
        \textbf{APFBM-GPO} & $\mathbf{50\%}$ & \textbf{Yes} & \textbf{Yes} \\
        \hline
    \end{tabular}
\end{table}

ACE extends outer constellation points to cancel peaks, incurring no SE loss but introducing signal distortion and providing no sensing capability. TR dedicates $N_{\text{TR}}$ subcarriers to peak-canceling tones, yielding an SE loss of $N_{\text{TR}}/N_{\text{data}}$ (typically $5$--$10\%$) with no sensing benefit. SLM/PTS require $\lceil\log_2(U)\rceil$ bits of SI per OFDM symbol, where $U$ is the number of candidate phase sequences; the SE loss is small but nonzero, and SI transmission failures degrade performance. By contrast, the proposed APFBM-GPO incurs a $50\%$ SE loss from the two-subcarrier-per-block mapping, yet simultaneously provides SI-free PAPR reduction \emph{and} pilotless sensing---a combination unavailable in any of the aforementioned schemes.

The SE cost of the APFBM scheme is therefore not solely a PAPR reduction overhead; it is the price of a signal structure that embeds both SI-free PAPR reduction and pilotless sensing within a single waveform~\cite{Xiong_ISAC_Limits_2023}. In ISAC deployments where dedicated sensing pilots or separate radar waveforms would otherwise consume spectral resources, the net overhead of APFBM may be comparable to the combined cost of conventional communication-plus-sensing designs, particularly when hardware efficiency and receiver simplicity are the primary constraints.

\section{Conclusion}
\label{sec:conclusion}

This paper has presented an APFBM scheme for OFDM-based ISAC that integrates SI-free PAPR reduction and pilotless sensing within a single waveform.
An unambiguous Stokes-to-Jones mapping establishes a deterministic phase reference per block, which the grouped GPO exploits to reduce PAPR without SI.
A non-iterative Viterbi-based receiver jointly recovers the GPO phases, demodulates data via GLRT with blind mismatch estimation, and performs DD-MRC channel estimation to reconstruct mismatch-aware CSI---all without dedicated sensing pilots.
Numerical results under a TGac Model-C channel demonstrate near-CRLB sensing accuracy and GPO recovery rates exceeding $99\%$ at moderate-to-high SNR.
Over-the-air experiments on an independent NI~X410 link at $5.4$~GHz further confirm practical decodability of $16$-APFBM and $64$-APFBM constellations, stable CSI recovery, and robust GPO performance.
The $50\%$ SE reduction inherent in the two-subcarrier-per-block mapping remains the primary limitation.
Future work will investigate multi-antenna extensions, higher-order APFBM constellations with improved SE, and tighter analytical bounds on the GPO recovery probability.

\appendices
\section{Derivation of the Rayleigh-Averaged SER}
\label{app:rayleigh_ser}
Let $h\sim\mathcal{CN}(0,\Omega)$ and $r\triangleq |h|$. The Rayleigh PDF is
\begin{equation}
    f_r(r)=\frac{2r}{\Omega}\exp\!\left(-\frac{r^2}{\Omega}\right),\quad r\ge 0.
\end{equation}
For $\beta>0$, the Rayleigh-averaged $Q$-function is $\mathbb{E}_{r}[Q(\beta r)] = \int_{0}^{\infty} Q(\beta r)\,f_r(r)\,dr$. Substituting Craig's representation of $Q(\cdot)$, exchanging the order of integration, evaluating the resulting Gaussian integral in~$r$, and applying the standard closed-form trigonometric identity yields
\begin{align}
\mathbb{E}_{r}[Q(\beta r)]
&=\frac{1}{2}\left(1-\frac{\beta\sqrt{\Omega}}{\sqrt{\beta^2\Omega+2}}\right).
\end{align}
Substituting $\beta\triangleq {d_{\min}}/{2\sqrt{2\sigma_w^2}}$ into \eqref{eq:ser_nn_bmisac} yields \eqref{eq:ser_nn_rayleigh_bmisac}.

\section{CRLB Derivation Under Multiplicative Intra-Pair Mismatch}
\label{app:crlb}
For notational simplicity, the block index $i$ is suppressed throughout this appendix; hence the scalar amplitude and phase parameters $(a,\theta)$ correspond to the per-block quantities $(a_i,\theta_i)$ in the main text, and $(x_a,x_b)$ denote the Jones components of one APFBM block. For the matched single-block model,
\begin{equation}
    \mathbf{z}=a e^{j\theta}\mathbf{x}+\mathbf{n},
    \qquad \mathbf{x}\triangleq [x_a,x_b]^T,
    \qquad \|\mathbf{x}\|_2^2=1.
    \label{eq:appendix_amp_linear_model}
\end{equation}
A first-order expansion at high SNR yields the amplitude perturbation $\hat{a}\approx a+\Re\!\left\{e^{-j\theta}\mathbf{x}^H\mathbf{n}\right\}$
and the phase perturbation (for $x_a\in\mathbb{R}_+$, $|n|\ll a|x_a|$)
\begin{equation}
    \arg(z)-\theta \approx \frac{\Im\!\left\{n e^{-j\theta}\right\}}{a|x_a|}.
    \label{eq:appendix_phase_imagpart_derivation}
\end{equation}

For the multiplicative-mismatch model,
\begin{equation}
    \mathbf{z} = a e^{j\theta}
    \begin{bmatrix}
        x_a \\
        \rho e^{j\varepsilon} x_b
    \end{bmatrix}
    + \mathbf{n},
    \label{eq:appendix_multiplicative_model}
\end{equation}
where $a\ge 0$, $\theta\in(-\pi,\pi]$, $\rho>0$, $\varepsilon\in(-\pi,\pi]$, $|x_a|^2+|x_b|^2=1$, and $\mathbf{n}\sim\mathcal{CN}(\mathbf{0},\sigma_w^2\mathbf{I}_2)$. The mean vector is
\begin{equation}
    \boldsymbol{\mu}(a,\theta,\rho,\varepsilon)=a e^{j\theta}
    \begin{bmatrix}
        x_a \\
        \rho e^{j\varepsilon} x_b
    \end{bmatrix}.
\end{equation}
Let $\boldsymbol{\eta}=[a,\theta,\rho,\varepsilon]^T$. The Fisher information matrix is
\begin{equation}
    [\mathbf{I}(\boldsymbol{\eta})]_{m,n}
    = \frac{2}{\sigma_w^2}\Re\!\left\{\left(\frac{\partial \boldsymbol{\mu}}{\partial \eta_m}\right)^H
    \left(\frac{\partial \boldsymbol{\mu}}{\partial \eta_n}\right)\right\}.
    \label{eq:fim_complex_gaussian_appendix}
\end{equation}
The required derivatives are
\begin{align}
    \frac{\partial \boldsymbol{\mu}}{\partial a}
    &= e^{j\theta}
    \begin{bmatrix}
        x_a \\
        \rho e^{j\varepsilon}x_b
    \end{bmatrix},
    &
    \frac{\partial \boldsymbol{\mu}}{\partial \theta}
    &= j a e^{j\theta}
    \begin{bmatrix}
        x_a \\
        \rho e^{j\varepsilon}x_b
    \end{bmatrix}, \\
    \frac{\partial \boldsymbol{\mu}}{\partial \rho}
    &= a e^{j\theta}
    \begin{bmatrix}
        0 \\
        e^{j\varepsilon}x_b
    \end{bmatrix},
    &
    \frac{\partial \boldsymbol{\mu}}{\partial \varepsilon}
    &= j a \rho e^{j\theta}
    \begin{bmatrix}
        0 \\
        e^{j\varepsilon}x_b
    \end{bmatrix}.
\end{align}
With $\alpha_x\triangleq |x_a|^2$ and $\beta_x\triangleq |x_b|^2$ ($\alpha_x+\beta_x=1$), substitution into \eqref{eq:fim_complex_gaussian_appendix} yields
\begin{equation}
    \mathbf{I}(\boldsymbol{\eta}) = \frac{2}{\sigma_w^2}
    \begin{bmatrix}
        \alpha_x+\rho^2\beta_x & 0 & a\rho\beta_x & 0 \\
        0 & a^2(\alpha_x+\rho^2\beta_x) & 0 & a^2\rho^2\beta_x \\
        a\rho\beta_x & 0 & a^2\beta_x & 0 \\
        0 & a^2\rho^2\beta_x & 0 & a^2\rho^2\beta_x
    \end{bmatrix}.
    \label{eq:fim_multiplicative_appendix}
\end{equation}
Partition \eqref{eq:fim_multiplicative_appendix} as
\begin{equation}
    \mathbf{I}(\boldsymbol{\eta})=
    \begin{bmatrix}
        \mathbf{I}_{11} & \mathbf{I}_{12} \\
        \mathbf{I}_{21} & \mathbf{I}_{22}
    \end{bmatrix},
\end{equation}
where $\mathbf{I}_{11}$ and $\mathbf{I}_{22}$ correspond to $(a,\theta)$ and $(\rho,\varepsilon)$, respectively. The Schur complement gives the equivalent Fisher information for $(a,\theta)$:
\begin{equation}
    \mathbf{I}_{\mathrm{eq}} = \mathbf{I}_{11} - \mathbf{I}_{12}\mathbf{I}_{22}^{-1}\mathbf{I}_{21}.
\end{equation}
This yields
\begin{equation}
    \mathbf{I}_{\mathrm{eq}} = \frac{2\alpha_x}{\sigma_w^2}
    \begin{bmatrix}
        1 & 0 \\
        0 & a^2
    \end{bmatrix},
    \label{eq:equivalent_fim_appendix}
\end{equation}
from which the CRLBs follow:
\begin{equation}
    \operatorname{Var}(\hat{a}) \ge \frac{\sigma_w^2}{2|x_a|^2},
    \qquad
    \operatorname{Var}(\hat{\theta}) \ge \frac{\sigma_w^2}{2a^2|x_a|^2}.
\end{equation}
Equivalently,
\begin{equation}
    \operatorname{RMSE}(\hat{a}) \ge \frac{\sigma_w}{\sqrt{2}|x_a|},
    \qquad
    \operatorname{RMSE}(\hat{\theta}) \ge \frac{\sigma_w}{\sqrt{2}a|x_a|}.
\end{equation}
These are the nuisance-aware CRLBs reported in \eqref{eq:crlb_amp_phase_block}.

When $\rho=1$ and $\varepsilon=0$, the model reduces to
\begin{equation}
    \mathbf{z}=a e^{j\theta}
    \begin{bmatrix}
        x_a \\
        x_b
    \end{bmatrix}+\mathbf{n},
\end{equation}
and the matched-model bounds become
\begin{equation}
    \operatorname{Var}(\hat{a}) \ge \frac{\sigma_w^2}{2},
    \qquad
    \operatorname{Var}(\hat{\theta}) \ge \frac{\sigma_w^2}{2a^2},
\end{equation}
or equivalently
\begin{equation}
    \operatorname{RMSE}(\hat{a}) \ge \frac{\sigma_w}{\sqrt{2}},
    \qquad
    \operatorname{RMSE}(\hat{\theta}) \ge \frac{\sigma_w}{\sqrt{2}a}.
\end{equation}
The gap between these matched-model bounds and the nuisance-aware bounds above quantifies the information loss incurred by the unknown mismatch parameters.

\section{Derivation of the Blind Amplitude-Mismatch Estimator}
\label{app:blind_rho}

This appendix derives the regression-based estimator \eqref{eq:roughness_beta} used in Step~2-1a. The block index $i$ is retained throughout.

From the GPO-compensated model \eqref{eq:gpo_compensated_block}, the squared magnitudes of the two components are
\begin{align}
    P_{a,i} &\triangleq |z_{a,i}|^2 = a_i^2 |\tilde{x}_{a,i}|^2 + w_{a,i}, \label{eq:app_pa}\\
    P_{b,i} &\triangleq |z_{b,i}|^2 = a_i^2 \rho_i^2 |\tilde{x}_{b,i}|^2 + w_{b,i}, \label{eq:app_pb}
\end{align}
where $w_{a,i}$ and $w_{b,i}$ collect the noise cross-terms and are $O(\sigma_w)$ at high SNR.

The key observation is that, for APFBM constellations, the energy-normalization constraint $|\tilde{x}_{a,i}|^2 + |\tilde{x}_{b,i}|^2 = 1$ holds per block, so the data-dependent terms $|\tilde{x}_{a,i}|^2$ and $|\tilde{x}_{b,i}|^2$ are complementary:
\begin{equation}
    |\tilde{x}_{b,i}|^2 = 1 - |\tilde{x}_{a,i}|^2.
    \label{eq:app_complement}
\end{equation}

Consider the first-order differences of the power sequences between adjacent blocks:
\begin{align}
    \Delta P_{a,i} &\triangleq P_{a,i+1} - P_{a,i}, \label{eq:app_dpa}\\
    \Delta P_{b,i} &\triangleq P_{b,i+1} - P_{b,i}. \label{eq:app_dpb}
\end{align}
Under two assumptions----(i)~the channel amplitude $a_i$ and mismatch magnitude $\rho_i$ vary slowly relative to the data symbols, so that $a_{i+1} \approx a_i \triangleq a$ and $\rho_{i+1} \approx \rho_i \triangleq \rho$ within each local window, and (ii)~the noise terms are negligible at high SNR---the differences are dominated by symbol-dependent fluctuations:
\begin{align}
    \Delta P_{a,i} &\approx a^2 \bigl(|\tilde{x}_{a,i+1}|^2 - |\tilde{x}_{a,i}|^2\bigr) \triangleq a^2 \, \delta_i, \label{eq:app_dpa_approx}\\
    \Delta P_{b,i} &\approx a^2 \rho^2 \bigl(|\tilde{x}_{b,i+1}|^2 - |\tilde{x}_{b,i}|^2\bigr). \label{eq:app_dpb_approx}
\end{align}
Applying the complementarity relation \eqref{eq:app_complement} to the second expression yields
\begin{equation}
    \Delta P_{b,i} \approx -a^2 \rho^2 \, \delta_i,
    \label{eq:app_dpb_neg}
\end{equation}
where $\delta_i \triangleq |\tilde{x}_{a,i+1}|^2 - |\tilde{x}_{a,i}|^2$. Within any local window, the two difference sequences are therefore related by a deterministic proportionality:
\begin{equation}
    \Delta P_{a,i} \approx -\frac{1}{\rho^2}\,\Delta P_{b,i}.
    \label{eq:app_linear_relation}
\end{equation}
This is a linear regression of $\Delta P_{a,i}$ on $\Delta P_{b,i}$ with slope $-1/\rho^2$ and zero intercept. Applying ordinary least-squares estimation over a local window $\mathcal{W}_i$ yields the slope estimate
\begin{equation}
    \hat{\beta}_i = -\frac{\sum_{j \in \mathcal{W}_i} \Delta P_{a,j}\,\Delta P_{b,j}}{\sum_{j \in \mathcal{W}_i} (\Delta P_{b,j})^2} \;\approx\; \frac{1}{\rho_i^2},
    \label{eq:app_beta_ols}
\end{equation}
from which $\hat{\rho}_i = 1/\sqrt{\hat{\beta}_i}$ follows immediately.

\ifCLASSOPTIONcaptionsoff
    \newpage
\fi

\bibliographystyle{IEEEtran}
\bibliography{Ref}

\end{document}